\documentclass[prb,twocolumn,showpacs,amsmath,amssymb,longbibliography]{revtex4-1}

\usepackage{bm}
\usepackage[dvipdfmx]{graphicx}
\usepackage{color}

\usepackage{mathdots} %imura 150429

\newcommand  {\eqn}[1]{(\ref{eqn:#1})}
\renewcommand{\(}     {\left(}
\renewcommand{\)}     {\right)}
\renewcommand{\[}     {\left[}
\renewcommand{\]}     {\right]}
\renewcommand{\_}[1]  {_{\rm #1}}

\begin{document}

\title{Dimensional crossover of transport characteristics in topological insulator nanofilms}
\author{Koji Kobayashi$^1$}
\author{Yukinori Yoshimura$^2$}
\author{Ken-Ichiro Imura$^{2,3}$}
\author{Tomi Ohtsuki$^1$}
\affiliation{$^1$Department of Physics, Sophia University, Tokyo Chiyoda-ku 102-8554, Japan}
\affiliation{$^2$Department of Quantum Matter, AdSM, Hiroshima University, Higashi-Hiroshima 739-8530, Japan}
\affiliation{
$^3$Kavli Institute for Theoretical Physics, University of California, Santa Barbara, CA 93106, USA
}

\date{\today}

\begin{abstract}
 We show how the two-dimensional (2D) topological insulator evolves, by stacking,
into a strong or weak topological insulator
with different topological indices,
proposing a new conjecture 
that goes beyond an intuitive picture
of the crossover from quantum spin Hall to the weak topological insulator.
 Studying the conductance
under different boundary conditions,
we demonstrate the existence of two conduction regimes
in which conduction happens
through either surface- or edge-conduction channels.
 We show that the two conduction regimes are 
complementary and exclusive.
 Conductance maps in the presence and absence of disorder are introduced,
together with 2D $\mathbb{Z}_2$-index maps,
describing the dimensional crossover of the conductance from the 2D to the 3D limit.
 Stacking layers is an effective way to invert the gap, 
an alternative to controlling the strength of spin-orbit coupling.
 The emerging quantum spin Hall insulator phase is not restricted to the case of odd numbers of layers.
\end{abstract}

\pacs{
%73.20.-r, % Electron states at surfaces and interfaces
73.20.At, % Electron states at surfaces and interfaces %%Surface states, band structure, electron density of states
%73.20.Fz, %  Weak or Anderson localization 
%73.21.Ac % 	Electron states and collective excitations in multilayers, quantum wells, mesoscopic, and nanoscale systems %%Multilayers
73.61.--r %Electrical properties of specific thin films
73.63.--b %Electronic transport in nanoscale materials and structures 
73.90.+f %Other topics in electronic structure and electrical properties of surfaces, interfaces, thin films, and low-dimensional structures
%71.23.-k, % Electronic structure of disordered solids
%71.30.+h  % Metal-insulator transitions and other electronic transitions
}
\maketitle

\section{Introduction} \label{sec:intro}

 Three-dimensional (3D) topological insulators (TIs) are classified into strong and weak,
\cite{MB,FKM,Roy}
depending on the number of gapless (Dirac) points in the surface Brillouin zone. 
 A strong topological insulator (STI) typically exhibits a single Dirac cone 
that is robust against disorder.
\cite{Shindou09, Goswami11, KOI, KOIH}
 The surface state of a weak topological insulator (WTI) with a pair of Dirac cones 
is superior in controllability,
e.g., its transport characteristics are sensitive to nanoscale formations of the sample.
\cite{yy}

%%%%%%%%%%%%%%%%%%%%%%%%%%
\begin{figure}[tbp]
\begin{tabular}{c}
\includegraphics[width=85mm, bb =0 0 464 202]{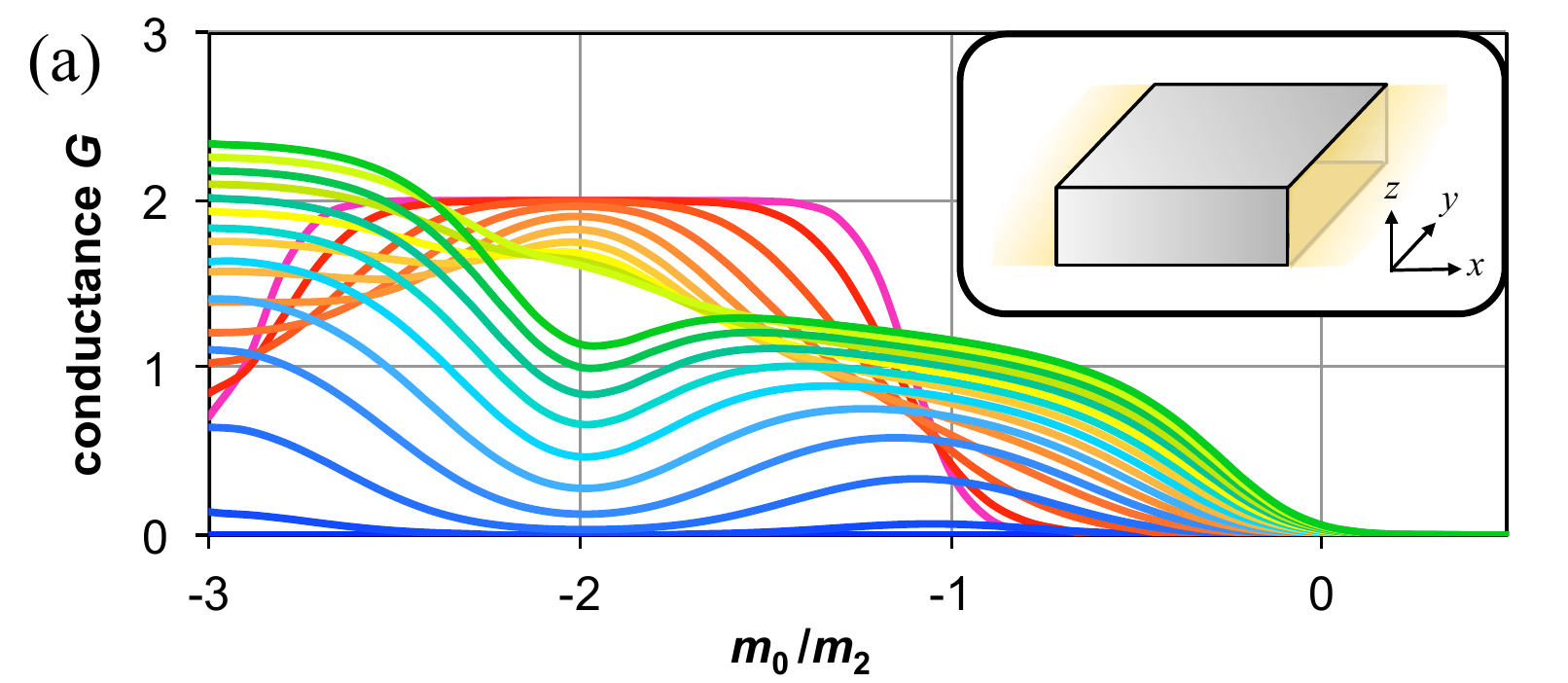}
\vspace{-0mm}
\\
\includegraphics[width=85mm, bb= 0 0 466 256]{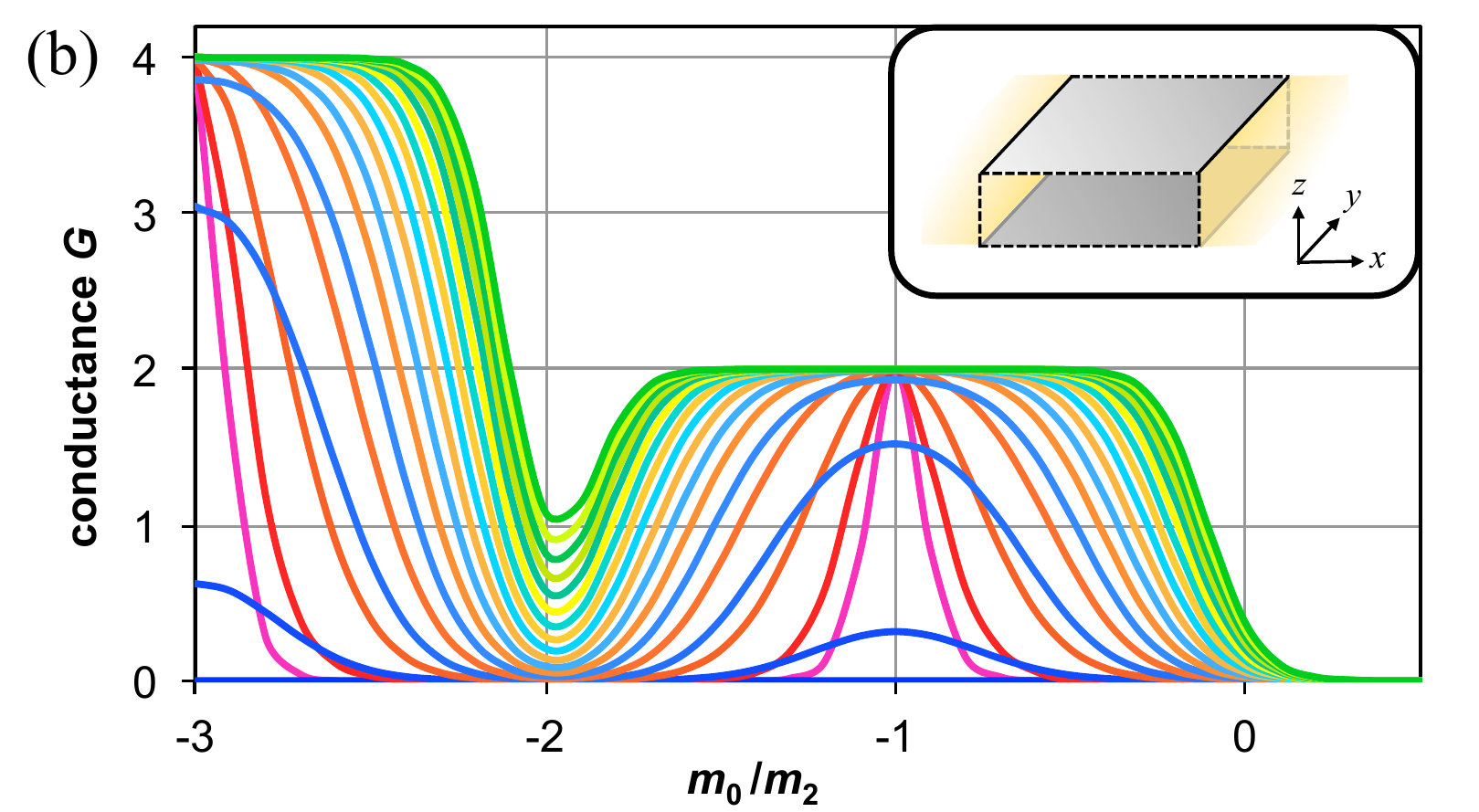}
\end{tabular}
\vspace{-2mm}
\caption{(Color online)
 Evolution of the conductance with variation of the film thickness $N_z$
and mass parameters ($m_0$ and $m_2$ are defined in Sec.~\ref{sec:model}).
 Different curves correspond to different $N_z$ values.
 Two types of boundary conditions
[side surfaces are (a) truncated vs (b) periodic]
are employed to highlight conduction plateaus of two different origins:
edge vs surface conduction.
 See Sec.~\ref{sec:conductance} for a detailed interpretation and settings.
}
\label{G-m}
\end{figure}
%%%%%%%%%%%%%%%%%%%%%%%

The existence of a topologically protected surface state
has been confirmed experimentally
by spin-resolved ARPES measurements
in a number of STI materials,\cite{Ando_JPSJ}
%which is now becoming a whole collection,
and is now firmly established.\cite{HasanKane}
 Recently, a WTI has also been experimentally identified.
\cite{WTI_exp}
Yet, as for triggering the topological nature of transport characteristics,
trials done in 3D bulk systems have been unsuccessful
due to the difficulty of separating the bulk and surface contributions 
in transport quantities.
\cite{Matsuo13}
 In this sense,
experimental observation of the topological nature of transport
is still restricted to two dimensions,
\cite{Molen}
and a natural step forward
is to extend this 2D result to the case of 3D TI nanofilms.
\cite{exp_film1, exp_film2, exp_film3, exp_film4,Bruene11}
 Indeed,
the study of TI nanofilms is of great interest recently
not only in its original but also in related contexts.
\cite{Fu, Scheurer14,Ozawa14,  Takane14,Liu15}

%%%%%%%%%%%%%%%%%%%%%%%%%%
\begin{figure*}[htbp]
\begin{tabular}{c}
\includegraphics[width=175mm, bb =0 0 740 220]{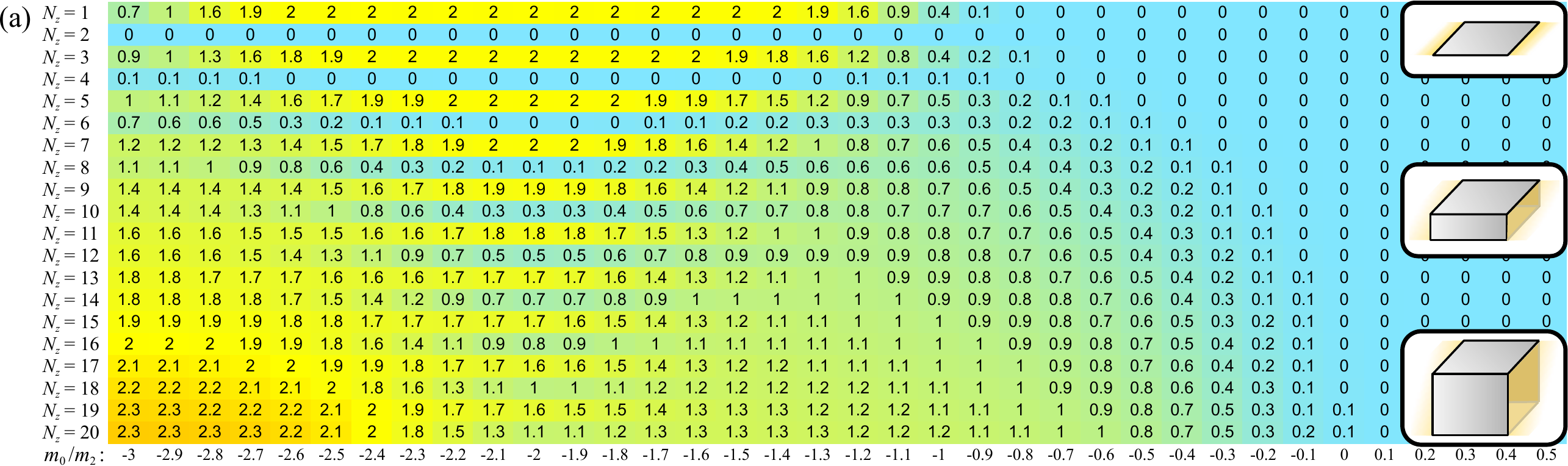}\\
\includegraphics[width=175mm, bb =0 0 741 220]{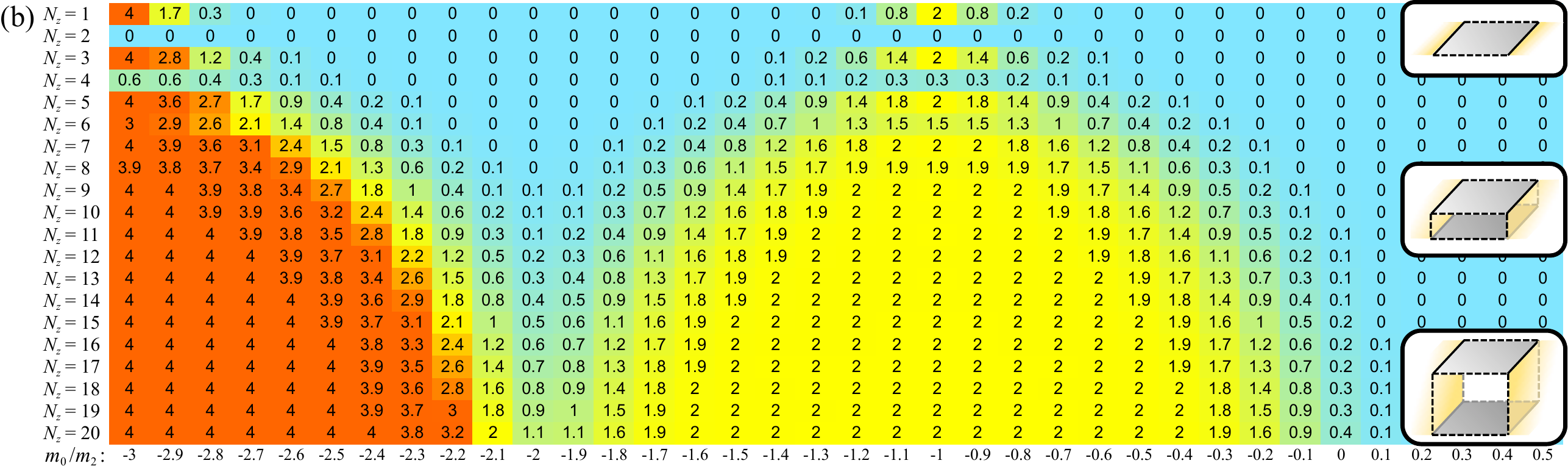}
\end{tabular}
\vspace{-2mm}
\caption{(Color online)
 The same conductance data as shown in Fig.~\ref{G-m}
are replotted in the form of ``conductance maps''
to highlight the edge and surface conduction regimes.
 Numbers listed are conductance values.
 See Sec.~\ref{sec:condMapDetail} 
for a detailed interpretation of the map.
}
\label{t20}
\end{figure*}
%%%%%%%%%%%%%%%%%%%%%%%

There exists a simple picture 
of the construction of a WTI
in which the WTI is viewed as stacked layers of quantum spin Hall (QSH) states,
i.e.,
``(QSH)$^{N_z} \simeq$ WTI''
in the limit of $N_z \rightarrow\infty$,
where $N_z$ is the number of stacked QSH layers.
\cite{TeoKane, Ringel, mayu1, Obuse14, Ozawa14, Takane14}
 This is indeed a useful point of view, explaining
how the helical edge state of a 2D QSH state 
evolves into the even number of Dirac cones 
on the surface of a WTI.
 Here, we examine
how precisely such a point of view
can be applied to the description of an actual crossover 
between the 2D and the 3D limits
in TI thin films.
\cite{yoko,Liu,Shen,ShenNJP,ebi,mayu2, Zhou14}
 An alternative view of the same dimensional crossover 
is to approach it from the 3D side,
i.e.,
starting with a bulk TI, and making the system thinner and thinner.
 In this second point of view,
evolution of the 3D topological features
is attributed to {\it hybridization} of the top and bottom surface wave functions
through the bulk.
 The first
``(QSH)$^{N_z} \simeq$ WTI''
picture is typically 
a surface (2D) point of view,
while the second picture is, in a sense,
a bulk (3D) point of view
and applicable in both the STI and 
the WTI regimes,
as long as the surface state wave function
extends to the top and bottom surfaces.
 In this paper we show,
by studying the 2D-3D crossover over a broad range of parameters,
how the above contrasting (surface- vs bulk-based) points of view
compensate each other to give a proper interpretation of the crossover phenomena
observed in different parameter regimes,
i.e., in different phases
(for example, STI vs WTI).
%, and WTIs with different weak indices).

 We examine this crossover from 
the two viewpoints.
 We first study the conductance of the nanofilm numerically
under different boundary conditions: 
truncated vs periodic
in the direction of the width of the film
(see Fig.~\ref{G-m}).
 Studying conductance
in these two different boundary conditions,
we demonstrate the existence of two conduction regimes:
{\it edge} and {\it surface} conduction regimes,
in which conduction happens
through either 
surface or edge conduction channels.
The two conductance maps obtained (Fig.~\ref{t20})
show that the two conduction regimes are 
complementary and exclusive.
 In order to interpret such conductance maps,
we then take the 
first (2D) viewpoint
in which TI thin film is regarded 
as an effective 2D system, 
and we calculate 2D $\mathbb{Z}_2$ indices,
introducing $\mathbb{Z}_2$-index maps.
 The $\mathbb{Z}_2$-index maps will demonstrate that
a 2D ordinary insulating layer is converted to a TI
as identical layers are stacked and coupled;
i.e., a nontrivial $\mathbb{Z}_2$ nature becomes {\it emergent} by stacking.

 The paper is organized as follows.
 In Sec.~\ref{sec:model} we introduce a standard model Hamiltonian
for the description of a bulk TI.
 Here, the same effective Hamiltonian is used to represent TI thin films
by implementing it as a tight-binding model.
 The main objective of the paper
is to interpolate different topological properties in the 2D and 3D limits.
 By its nature, the 2D or 3D $\mathbb{Z}_2$ index alone
is insufficient for the description of this dimensional crossover.
 In Sec.~\ref{sec:condmap},
we overcome this difficulty
by studying conductances of TI nanofilms
and establish conductance maps as shown in Fig.~\ref{t20}.
 A sketch of the employed methods and
interpretation of the obtained conductance maps are given there.
 Then in Sec.~\ref{sec:Z2map},
we attempt to classify such TI thin films of varying layer thicknesses
in terms of the 2D $\mathbb{Z}_2$ indices.
 We establish the phase diagrams
of the thin films as the effective 2D system.
 Effects of disorder are addressed in Sec.~\ref{sec:disorder},
before the paper is concluded in Sec.~\ref{sec:conclusion}.

%%%%% II. Model %%%%%

\section{Model Hamiltonian} \label{sec:model}
 Let us start by defining our model Hamiltonian and its parameters.
 The model Hamiltonian is introduced in the 3D, bulk limit.
 It turns out to be convenient to define the bulk Hamiltonian on a lattice,
i.e., as a tight-binding Hamiltonian.
 Here, the type of this lattice is chosen to be {\it cubic}, 
and the results of all subsequent analyses
are superficially dependent on the choice of this lattice.
 Then, by truncating this tight-binding Hamiltonian
in real space we model a TI nanofilm.

%\section{Thin film model and $\mathbb Z_2$ index map}
\subsection{Bulk: 3D limit}

 As a model for a 3D bulk TI,
we consider the following Wilson-Dirac model,
here implemented as a tight-binding Hamiltonian
on a cubic lattice,
%---
\begin{align}
 H_{\rm bulk}(\bm k)  = 
 m(\bm k)\beta + \sum_{\mu=x,y,z} t_\mu\sin k_\mu \alpha_\mu,
\label{H_bulk}
\end{align}
%---
where 
%---
\begin{align}
m(\bm k) = m_0 + \sum_{\mu=x,y,z} m_{2\mu} (1-\cos k_\mu).
\label{m_bulk}
\end{align}
%---
 The length unit is set to the lattice constant.
 $H(\bm k)$ is a $4\times 4$ matrix,
spanned by two types of Pauli matrices $\bm \sigma$ and $\bm \tau$,
each representing physically real and orbital spins.
 The explicit form of the Wilson-Dirac Hamiltonian, 
Eq.~(\ref{H_bulk}),
corresponds to the choice of $\alpha$ and $\beta$ matrices such that
%___ alpha beta ___
\begin{align}
 \alpha_\mu = \tau_x \otimes \sigma_\mu,\ \ 
 \beta   = \tau_z \otimes 1_2,
\label{gamma}
\end{align}
%---
where $\mu=x,y,z$.
The ``Dirac nature'' of 
the Hamiltonian
is encoded in the anti-commutation relation of these matrices:
%---
\begin{align}
 \{ \alpha_\mu, \alpha_{\mu'} \}
 = 2 \delta_{\mu{\mu'}},  \ \ 
 \{ \alpha_\mu, \beta \}
 = 0.
\label{anticom}
\end{align}
%---

\subsection{Thin film: An effective 2D system}
 The tight-binding form of the bulk effective Hamiltonian, 
Eq.~(\ref{H_bulk}),
is useful for implementing the thin film geometry.
 To realize a film of thickness $N_z$
extended in the $(x,y)$ plane,
we first rewrite the $k_z$ dependence of Eqs.~(\ref{H_bulk}) and (\ref{m_bulk})
in terms of real-space hopping terms,
then truncate the hopping outside the system composed of $N_z$ layers
starting at $z=1$ and ending at $z=N_z$.
 The explicit form of such a tight-binding Hamiltonian reads
%___ H_film ___
\begin{align}
 H_{\rm film}(k_x,k_y) \!=\!
  &\ 
   1_{N_z}\!\otimes \!\!
   \(
     m_{\rm 2D}^{(k_x,k_y)} \beta
    \!+\!\!\!\! \sum_{\mu=x,y}\!\! t_\mu\sin k_\mu \alpha_\mu
   \) 
%   \nonumber \\
%  &
   \!+\!H_z ,
\label{H_film}
\end{align}
%---
where
%___ m_film ___
\begin{align}
 m_{\rm 2D}^{(k_x,k_y)} = &
   \(
    m_0 + m_{2z} +\sum_{\mu=x,y} m_{2\mu} (1-\cos k_\mu)
   \),
\end{align}
%---
and
%___ m_film ___
\begin{align}
 H_{z} = &
   \begin{pmatrix}
    0 & 1      &        &   \\[-2.0mm]
    1 & \ddots & \ddots &   \\[-1.5mm]
      & \ddots & \ddots & 1 \\[-0.0mm]
      &        & 1      & 0
   \end{pmatrix}
   \otimes    \left(-{m_{2z}\over 2} \beta \right) 
   \nonumber \\
   +&
   \begin{pmatrix}
    0 & -1     &        &   \\[-2.0mm]
    1 & \ddots & \ddots &   \\[-1.5mm]
      & \ddots & \ddots & -1\\[-0.0mm]
      &        & 1      & 0
   \end{pmatrix}
   \otimes   \left({\rm i}{t_{z}\over 2} \alpha_z \right) .
\label{H_z}
\end{align}
%---
Here, the first part of the direct products in Eqs.~(\ref{H_film}) and (\ref{H_z})
represent the layer degrees of freedom.
This Hamiltonian
consists of two types of contributions;
the first term
of Eq.~(\ref{H_film})---we may call it $H_{xy}$---is 
composed of diagonal blocks, 
each corresponding to a different layer
and consisting of an effective 2D Hamiltonian in the layer,
while the second term,
$H_z$, represents hopping between such layers.

%%% real space  %%%
\subsection{Nanofilm case: Study of the dimensional crossover}

 Since the main interest in this paper is the study of a dimensional crossover
between 2D and 3D limits, we employ TI nanofilms 
with their boundaries truncated in all three spatial directions, 
i.e., finite-sized samples.
% Since main interest of this paper is the dimensional crossover case between 2D and 3D limits,
%we employ TI nanofilms with truncated boundaries, i.e., finite sized samples.
 Considering TI nanofilms,
we found it useful to
rewrite Hamiltonian (\ref{H_bulk}) in real space as
%___ eqn:H_real ___%
   \begin{align}
      H &=  \sum_{\bf x} \sum_{\mu=x,y,z} \(\frac{{\rm i}t_\mu}{2} c^{\dag}_{{\bf x}+{\bf e}_\mu} \alpha_{\mu} c_{\bf x}
                                         -{m_{2\mu}\over 2}  c^{\dag}_{{\bf x}+{\bf e}_\mu} \beta c_{\bf x} + \rm{h.c.}\)  \nonumber \\
            & + (m_0+ m_{2x}+ m_{2y}+ m_{2z})\sum_{\bf x} c^{\dag}_{\bf x} \beta c_{\bf x},
 \label{eqn:H_real}
\end{align}
%---%
where ${\bf x}$ is a position of lattice sites and ${\bf e}_\mu$ is the lattice vector in the $\mu$ direction.
%\textcolor{red}{%vvv red vvv
%We measure the size of the sample in terms of the number of sites, as $L_x \times L_y \times N_z$.
%}%^^^ red ^^^

%%%%%%%%%%%%%%%%%%%%%%%%%%
\begin{figure}[tbp]
\begin{tabular}{r}
\includegraphics[width=85mm, bb =0 0 693 393]{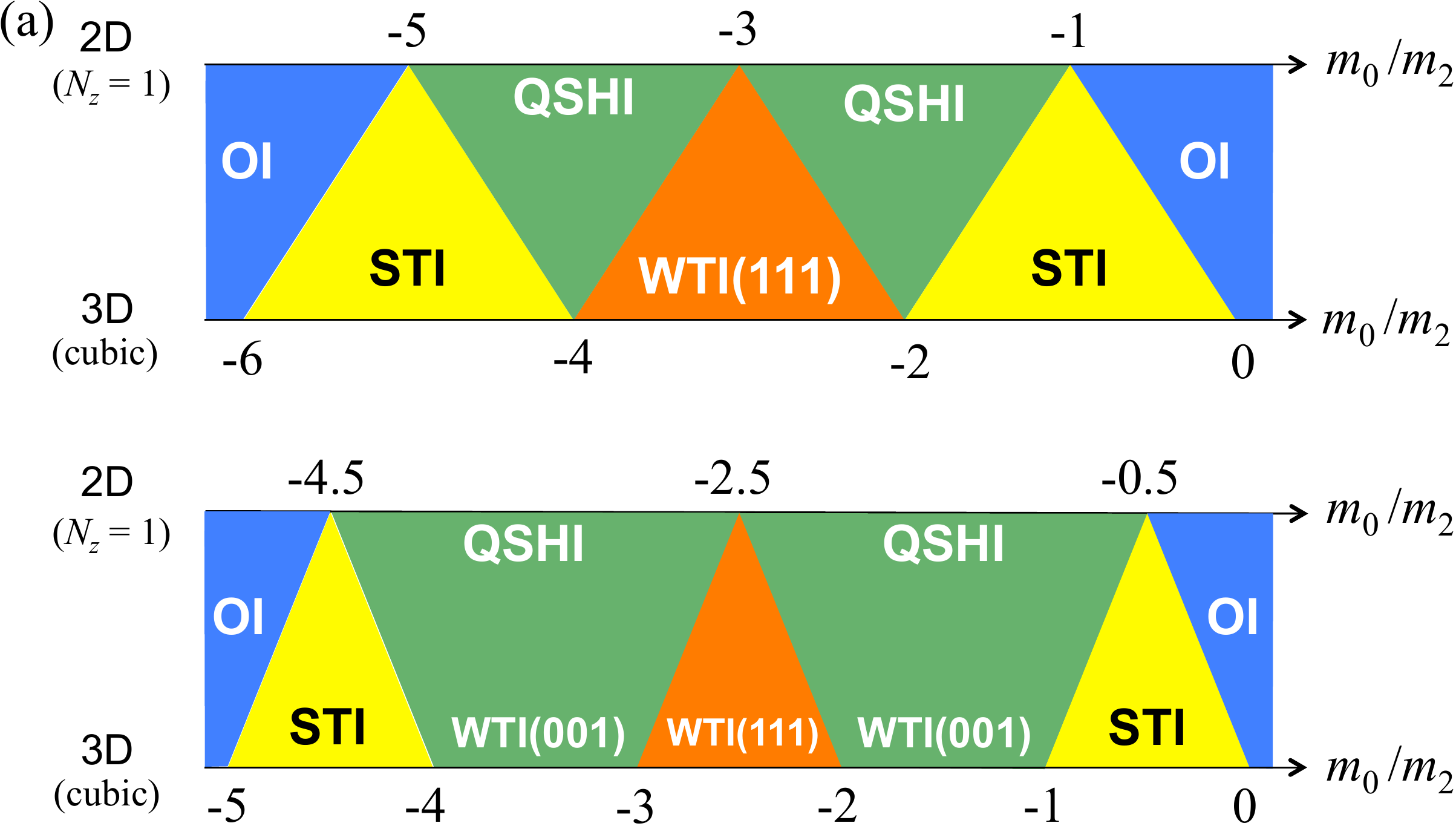}\\[5mm]
\includegraphics[width=85mm, bb =0 0 693 393]{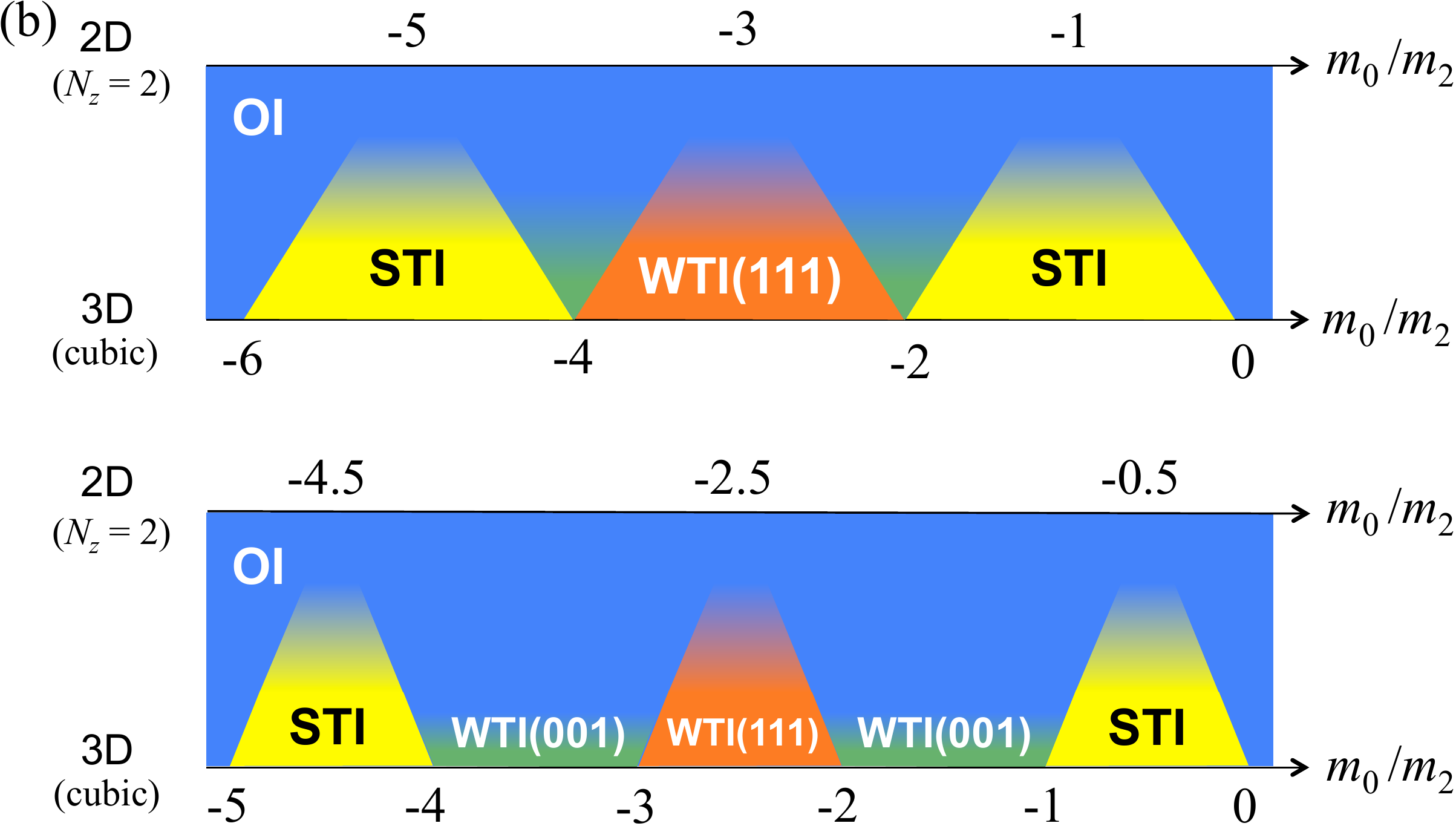}
\end{tabular}
\vspace{-2mm}
\caption{(Color online)
Phase diagram: an abstracted form of the conductance map.
Comparison of the cases of (a) $N_z$ odd and (b) $N_z$ even.
Comparison of the isotropic (upper panel) vs anisotropic (lower panel) cases; see Eqs.~\eqn{iso} and \eqn{aniso}.
Valid in the regime of $|t_z/m_{2z}|>1$.
The edge (surface) conduction regimes are represented as 
green (yellow/red) areas.
}
\label{PD}
\end{figure}
%%%%%%%%%%%%%%%%%%%%%%%

%%%%% conductance map %%%%%

\section{Conductance map}
\label{sec:condmap}

Using the model Hamiltonian, 
Eq.~\eqn{H_real}, 
introduced in the previous section,
we study the conductance of TI nanofilms numerically.
 Studying the conductance
under different boundary conditions (recall Fig.~\ref{G-m}),
we demonstrate the existence of two conduction regimes:
the {\it edge and surface conduction} regimes.
 To reveal the nature of these two conduction regimes,
especially with regard to its thickness dependence,
%especially the thickness dependence,
the same data as shown in Fig.~\ref{G-m}
are replotted in Fig.~\ref{t20}
in a different format:
as conductance maps.

%%%%%  %%%%%
\subsection{Numerical study of conductance} \label{sec:conductance}

 We focus on
the two terminal conductance $G$ of TI nanofilms,
hereafter measured in units of $e^2/h$,
and at Fermi energy $E=0$.
 Using the transfer matrix method,\cite{KOI}
we calculate the transmission coefficients for each conduction channel,
then deduce the two-terminal conductance $G$
with the help of the Landauer formula.

 To study the behavior of $G$,
we employ the geometry shown in the insets in Fig.~\ref{G-m}.
 We also employ two types of boundary conditions,
truncated vs periodic,
in the direction of the width of the film ($y$ direction),
corresponding to a film with vs without edges, respectively.
 Here, we have placed the sample in such a way that
the film is on the $xy$ plane, while leads are attached to the two extremities
of the sample in the $x$ direction;
i.e., the conduction occurs in the $x$ direction.
 The leads consist of 1D perfectly conducting channels, so that the effect of the leads is minimized.

 Figure~\ref{G-m} shows typical examples of our conductance data.
 In both panels,
the two-terminal conductance $G$ is plotted
as a function of $m_0/m_2$;
i.e.,
the evolution of the conductance 
with the change of 
the mass parameter $m_0$
introduced in Eq.~(\ref{m_bulk}).
 Here, the mass parameters $m_{2x}$, $m_{2y}$ and $m_{2z}$ are chosen to be isotropic:
%___ m_2x=m_2y=m_2z ___%
\begin{align} \label{eqn:iso}
   m_{2x}=m_{2y}=m_{2z}=m_2.
\end{align}
%---%
 We have also set the strength of hopping $t_z$
such that $t_z/m_{2z}=2$.
 Different curves in Fig.~\ref{G-m} correspond to different thicknesses $N_z$ of the film.
The size of the system is such that
the top and bottom surfaces are of size
$L_x\times L_y = 20 \times 20$,
while the thickness $N_z$ of the film is varied from 1 to 20.  
Reddish curves, to the case of $N_z$ odd;
bluish curves correspond to the case of $N_z$ even;
while greenish curves, to cases where $N_z$ is so large that the system looks like a cubic geometry rather than a film.

 In the 2D limit,
i.e.,
at $N_z=1$, 
the system is either in the QSH phase 
[$\nu^{\rm 2D}_0=1$,
see Eq.~\eqn{nu_0} for definition]
with a pair of helical edge modes,
or in the ordinary insulator (OI) phase 
($\nu^{\rm 2D}_0=0$)
without an edge mode,
depending on the value of $m_0/m_2$.
 The QSH phase occurs in the range $-3<m_0/m_2<-1$,
while the OI phase corresponds to $-1<m_0/m_2$.
 The QSH phase implies a plateau of the conductance at $G=2$
in the truncated geometry [Fig.~\ref{G-m}(a)],
while $G=0$ is expected in the OI phase.
 Such a tendency can be seen in the reddish curves 
in Fig.~\ref{G-m}(a).
 Note that in this truncated geometry
the contribution from the edge conduction channels is accentuated.

However,
in the 3D limit
%at $N_z\rightarrow \infty$, 
or, more practically, 
in the cubic geometry $N_z=L_x=L_y$, 
%as large as an isotropic value,
the phase boundaries between topologically different phases 
are shifted from the above 2D values.
 The system is expected to be
in the STI phase in the range $-2<m_0/m_2<0$,
while it is in the WTI phase with weak indices $\bm\nu^{\rm 3D}= (111)$ at $-4<m_0/m_2<-2$.
 The OI phase corresponds to $0<m_0/m_2$.
 Let us focus on Fig.~\ref{G-m}(b),
the case of periodic boundary conditions.
 In this geometry, the film has no edge so that
only contributions from the top and bottom surfaces
are well maintained, 
as long as the bulk is gapped 
and insulating.
 In the STI phase
the existence of a single Dirac cone on the top and bottom surfaces
implies a conductance plateau at $G=2$,
while in the WTI phase
double Dirac cones on the top and bottom surfaces sum up to $G=4$.
 One can see a precursor of such plateaus in the greenish curves in Fig.~\ref{G-m}(b).

%%% CM %%%
\subsection{The conductance map} \label{sec:condMapDetail}
 We have so far seen that the Figs.~\ref{G-m}(a) and \ref{G-m}(b)
indeed show conductance plateaus of two different natures:
the edge and surface conduction types.
 They also appear in different parameter regimes,
so that there are edge and surface conduction regimes.
 In Fig.~\ref{t20}
we have replotted the same data as shown in Fig.~\ref{G-m}
in a different format,
as ``conductance maps'',
to reveal the nature of these two conduction regimes.
 Presented in this way,
the existence of two conduction regimes
may become apparent.

%\subsubsection{Crossover from 2D to 3D}

\subsubsection{Edge vs surface conduction regimes}

 The nature of the two conductance plateaus is apparent in Figs.~\ref{t20}(a) and \ref{t20}(b).
 The conductance plateaus at $G=2$
associated with the edge conduction channels
that appear in the regime of $m_0/m_2$ centered at $m_0/m_2=-2$ in Fig.~\ref{G-m}(a)
are arranged into the vertically striped, downward-pointing triangle region
at the left in Fig.~\ref{t20}(a): 
the {\it edge conduction regime}.
 Such conductance plateaus are shown by the reddish curves in Fig.~\ref{G-m}(a),
i.e., at an odd number of layers,
while the bluish curves, 
i.e., at an even number of layers,
show vanishing conductance in the same range of $m_0/m_2$;
therefore, an alternating pattern of $G=2$ and $G=0$ is shown in the conductance map in Fig.~\ref{t20}(a).
 Note that in the conductance map
the vertical axis represents the film thickness $N_z$,
encoding the dimensional crossover of transport characteristics.

 Similarly,
the conductance plateaus at $G=2$ and at $G=4$ in Fig.~\ref{G-m}(b)
form the two upward-pointing triangular regions in Fig.~\ref{t20}(b),
corresponding to STI and WTI regimes.
 Naturally,
such conduction plateaus are associated with bulk conduction channels,
so they can be seen as the greenish curves in Fig.~\ref{G-m}(b).
 In Fig.~\ref{t20}(b),
the two corresponding triangular regions 
become predominant in the limit of large $N_z$.
 As $N_z$ is decreased,
a region of vanishing conductance appears 
between the two triangular regions,
which corresponds, in the parameter space of $(m_0/m_2, N_z)$,
to the edge-conduction 
regime 
in Fig.~\ref{t20}(a).

 Quantized conduction in TI nanofilms
is caused by either the surface, or the edge conducting channels.
 The two types of conducting channels 
are not 
completely open at the same 
time.
Only one 
of the channels is always open
except in the crossover regime,
in which both channels are partially open.
 The two conduction regimes are, therefore,
both {\it complementary} and {\it exclusive}.
 Away from the 2D or 3D limits,
at which transport characteristics are simply determined by
the 2D or 3D $\mathbb Z_2$ topological indices,
the conductance maps cast a new perspective on the competitive nature of 
2D TI-like (edge conduction) and 3D TI-like (surface conduction) properties.

%%%  %%%
\subsubsection{Surface point of view: Approach from the 2D limit}

 In the 2D limit, i.e., at $N_z=1$, 
the QSH phase occurs in the range of parameters $-3<m_0/m_2<-1$.
 A plateau of the quantized conductance at $G=2$ appears
in this parameter regime,
due to the pair of helical edge modes 
circulating around the gapped 2D bulk area.
 Stacking such a layer in the $z$ direction,
one can constract a TI nanofilm with helical edge modes circulating 
around the side surfaces.
 When $N_z$ layers are stacked,
$N_z$ pairs of helical edge modes are incident on the side surfaces
and couple with each other.
 If $N_z$ is even, they are all gapped, while
if $N_z$ is odd, there always exists a single combination that remains gapless
and perfectly conducting even in the presence of disorder.
 The origin of the striped pattern in the edge-conduction regime in Fig.~\ref{t20}(a)
is this even/odd feature.

 In the above picture
the helical edge modes circulating around the side surfaces
are expected to become an even number of Dirac cones
characteristic of a WTI surface,
while
the top and bottom surfaces are kept gapped.
 This situation is realized in a WTI state with weak indices $\bm\nu^{\rm 3D}= (001)$,
and this indeed happens in the case of the standard ``(QSH)$^{N_z} \simeq$ WTI'' picture.
 However,
the situation we have here is slightly different.
 In the 3D limit of our system,
a WTI phase is indeed expected in the range of parameters $-4<m_0/m_2<-2$,
but its weak indices are $\bm\nu^{\rm 3D}= (111)$;
i.e.,
the top and the bottom surfaces are {\it not} gapped in this phase.
 Also, an STI phase is expected in the range of parameters $-2<m_0/m_2<0$.
 In the 2D-3D crossover shown in the conductance maps in Fig.~\ref{t20},
there is no WTI state
expected from the 2D side
as a consequence of the standard ``(QSH)$^{N_z} \simeq$ WTI'' crossover.
 In Fig.~\ref{t20}(a)
the region showing a striped pattern
indeed becomes narrower as $N_z$ is increased toward the line of $m_0/m_2=-2$.

 What, then, happens in our 2D-3D crossover regime?
 Actually,
what can be regarded as
a WTI state with weak indices $\bm\nu^{\rm 3D}= (001)$
is uncoupled layers of QSH states stacked in the $z$ direction,
which corresponds in the present context to the limit of
%___ m_perp=0 ___%
\begin{align}
   m_{2\perp}=m_{2z}=0.
\end{align}
%---%
 The phase diagram of our model Hamiltonian, (\ref{H_bulk}),
in the 3D limit
in the case of such an anisotropic choice of parameters
can be found in Fig.~1 in Ref.~\onlinecite{mayu1},
where
%___ m_2x=m_2y ___%
\begin{align}
   m_{2\parallel}=m_{2x}=m_{2y}.
\end{align}
%---%
 One notes in that phase diagram that
the $m_{2\perp}=0$ line in this 3D phase diagram is nothing but
the 2D limit we have already considered;
only the QSH phase is replaced by 
a WTI state with weak indices $\bm\nu^{\rm 3D}= (001)$.
 Starting with this limit and switching on $m_{2\perp}$ adiabatically,
one can follow the evolution of the system 
from this 2D limit to an isotropic line, $m_{2\perp}=m_{2\parallel}$, 
which we regard here as the 3D limit.
 The STI phase and 
the WTI phase with weak indices $\bm\nu^{\rm 3D}= (111)$
emerge and 
replace the WTI (001) phase 
%state with weak indices $\bm\nu^{\rm 3D}= (001)$
as $m_{2\perp}/m_{2\parallel}$ is switched on and approaches unity.
 A similar phase diagram,
consisting of cross-dimensional topological regions,
has been proposed recently
in the context of cold atoms.
\cite{Scheurer14}
 This is indeed, at least theoretically, a reasonable way to achieve 
the 2D-3D crossover.
 Yet, ``physically'' a more natural way to induce that crossover is 
to stack layers.
 Stacking layers is, 
however, not a continuous deformation, 
since the number of layers can be changed only discretely.
 A number of different topological phases emerge 
as an increasing number of layers is stacked.
 As the even/odd feature is washed out 
with increasing $N_z$ in Fig.~\ref{t20}(a),
quantized conductance appears 
as a new emergent feature
in the two triangular regions in Fig.~\ref{t20}(b), 
i.e.,
in the surface-conduction regimes.

 We performed a conductance study
also in such an anisotropic regime of parameters, 
and here we show the results only schematically.
%but here shows its results only schematically. 
 In Fig.~\ref{PD}
we divide the information from the conductance map 
into cases of $N_z$ odd [Fig.~\ref{PD}(a)] and $N_z$ even [Fig.~\ref{PD}(b)], 
then
abstract it into the form of a schematic phase diagram.
 The upper panels in Figs.~\ref{PD}(a) and \ref{PD}(b)
correspond to the case of an isotropic choice of parameters, 
Eq.~\eqn{iso},
while
the lower panels represent the anisotropic case:
%___ m2perp = m2/2 ___
\begin{align}
   m_{2\perp}=m_{2\parallel}/2.
 \label{eqn:aniso}
\end{align}
%---
 In the figure $m_{2\parallel}$
is simply represented as $m_2$.
 On the anisotropic line, 
Eq.~\eqn{aniso},
there indeed exists a range of parameters, $-2<m_0/m_2<-1$,
in which ``(QSH)$^{N_z}$'' leads to an expected WTI with $\bm{\nu}^{\rm 3D}=(001)$.

%%% 3D limit %%%
\subsubsection{Bulk point of view : Approach from the 3D limit} \label{sec:3Dlimit}

%We have so far viewed the conductance map from the 2D, QSH side. Here, we attempt to give an alternative view from the 3D limit.
 The film geometry of an STI in the
shows ``pseudo gapless'' surface states in both the top and the bottom surfaces of the film.
 Here,
what we imply by the quotation marks is
that usually the surface states are slightly gapped due to hybridization between the top and the bottom surface states.
 The corresponding wave function has an exponentially decaying tail into the bulk,
and the magnitude of the hybridization gap\cite{mayu2} is essentially determined by
the remaining amplitude at the opposite surface,
i.e., $\psi_{z=N_z}$ 
of the top surface wave function.
 As a result, surface conduction is suppressed with decreasing $N_z$,
while
in the cubic (3D) case on the bottom low in Fig.~\ref{t20}(b),
the hybridization between the top and the bottom surfaces 
can be neglected.

 In the 3D limit, the STI phase is extended over $-2<m_0/m_2<0$.
 The penetration of such a surface state
becomes deeper as one approaches the 3D topological phase boundary 
at $m_0/m_2=0,-2$ (where the Dirac semimetal phase appears).
 Near the 3D phase boundary,
as one reduces the number of layers $N_z$,
a hybridization gap is formed more easily,
driving the system to the edge conduction regime at relatively large $N_z$.
 With decreasing $N_z$,
quantization of the conductance due to the surface Dirac cones
is naturally destroyed, while
at the center of the STI phase at $m_0/m_2=-1$
the surface conduction survives exceptionally
down to the single-layer (2D) geometry: $N_z=1$.
 This is because at $m_0/m_2=-1$,
the matrix elements between surface states on the top and bottom surfaces vanish
and the hybridization gap does not open 
as long as $N_z$ is odd.

%%%%%%%%%%%%%%%%%%%%%%%%%%
\begin{figure}[tbp]
\begin{tabular}{r}
\includegraphics[width=85mm, bb= 0 0 427 238]{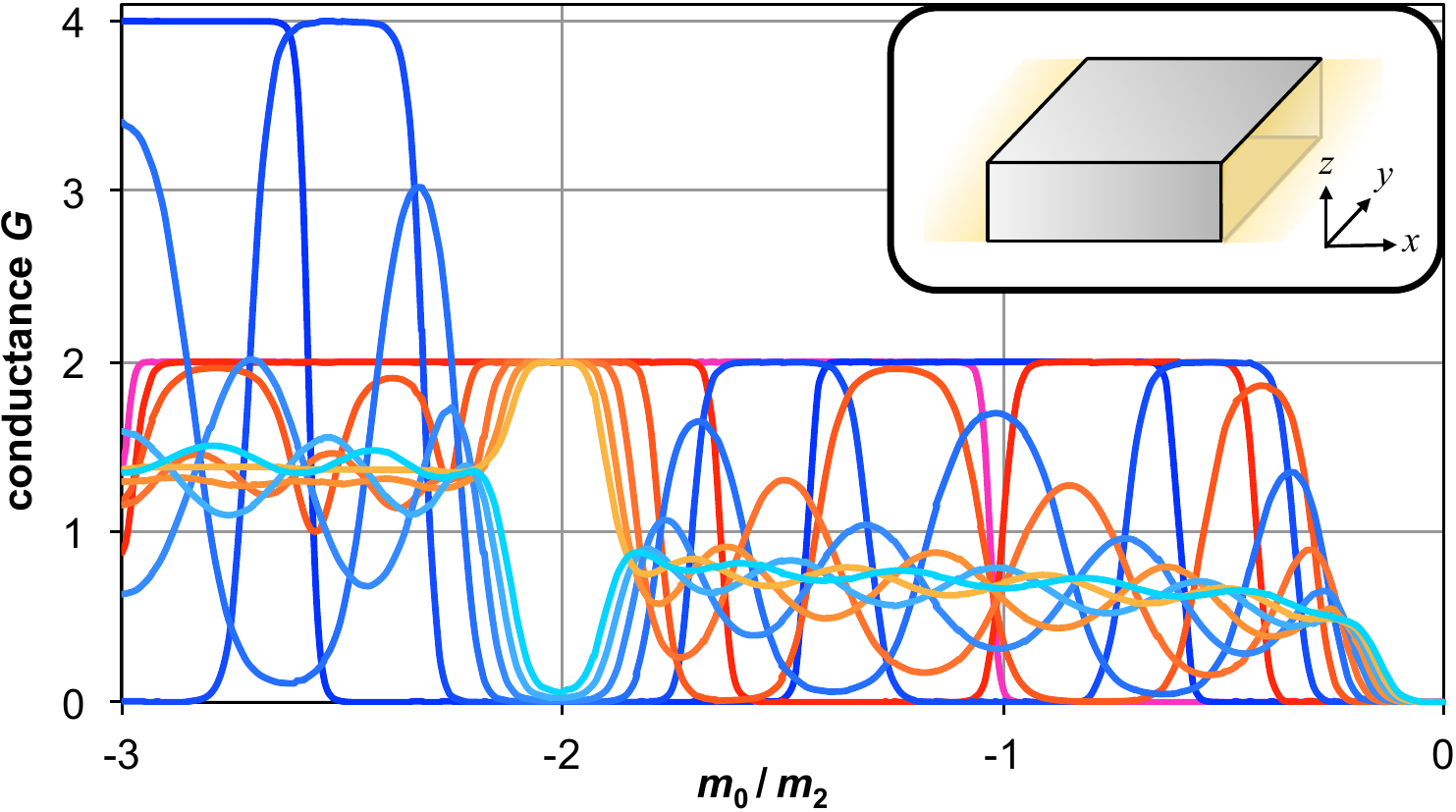}
\end{tabular}
\vspace{-2mm}
\caption{(Color online)
 Conductance of TI nanofilms as in Fig.~\ref{G-m}(a), 
but with $t_z/m_{2z}=0.5$.
 In contrast to Fig.~\ref{G-m}(a), the behavior of the conductance looks rather erratic.
 The corresponding conductance map is given in Fig.~\ref{t05}(b).
}
\label{G-m_zebra}
\end{figure}
%%%%%%%%%%%%%%%%%%%%%%%

%%%%%%%%%%%%%%%%%%%%%%%%%%
\begin{figure*}[tbp]
\begin{tabular}{c}
\includegraphics[width=178mm, bb =0 0 754 157]{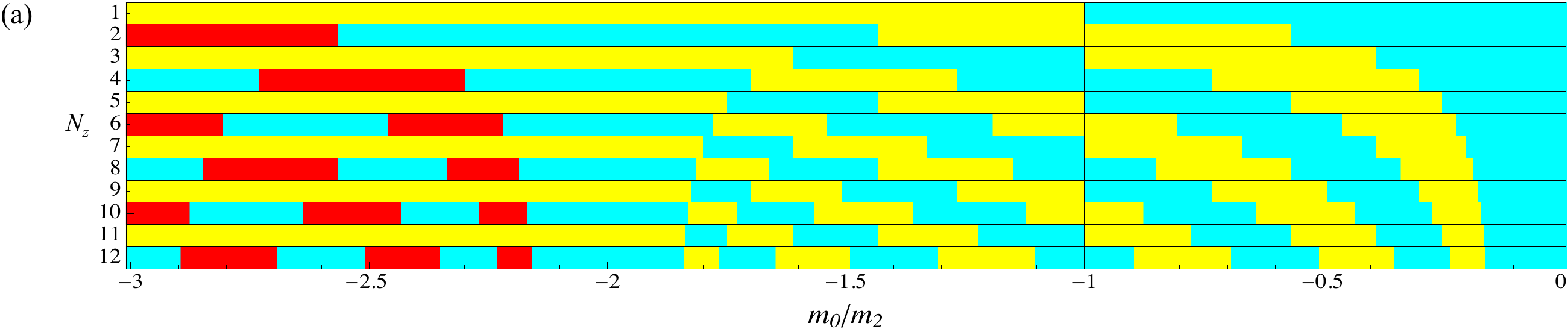}
\\[2mm]
\includegraphics[width=177mm, bb= 0 0 753 139]{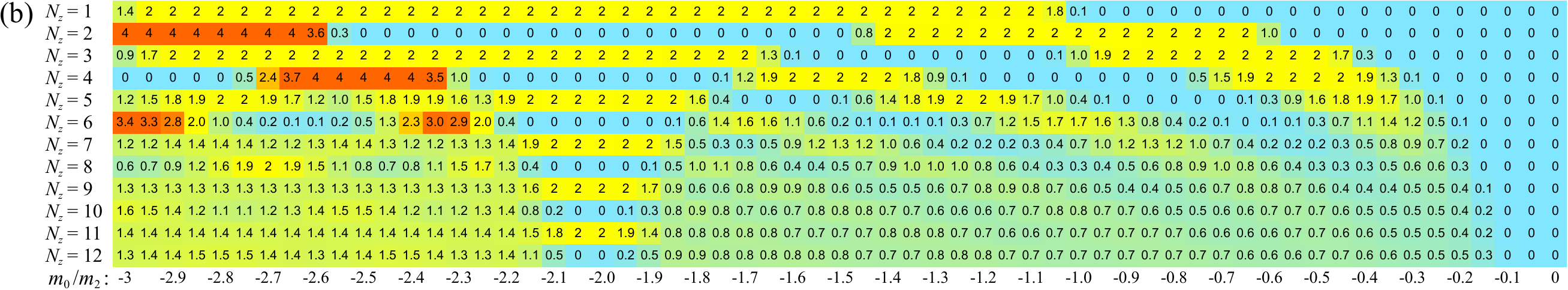}
\end{tabular}
\vspace{-2mm}
\caption{(Color online) 
(a) $\mathbb{Z}_2$-index map vs (b) conductance map 
in the case of a relatively small $t_z$, $t_z/m_{2z}=0.5$;
cf.
$t_z/m_{2z}=2$ 
in the regime of $t_z > 1$
in Fig.~\ref{t20}.
 In the regime of relatively small $t_z$,
a stripe pattern prominent in the lower-triangular region
in Fig.~\ref{t20}(a)
(the edge conduction regime) is extended to a broader range of parameters
on the side of smaller $|m_0/m_2|$.
 But outside the original triangular regime
this is no longer an even/odd feature:
an effective QSH insulator (QSHI), or 2D TI ($\nu^{\rm 2D}_0=1$) phase (yellow region)
appears not just with an odd number of layers.
 The trivial phase with $\nu^{\rm 2D}_0=0$ corresponds to the blue region,
while
the red region corresponds to a 2D analog of the more standard 3D weak TI phases.}
\label{t05}
\end{figure*}
%%%%%%%%%%%%%%%%%%%%%%%

%%% break %%%
\subsubsection{Breakdown of the even/odd feature}

 Figure~\ref{G-m_zebra} shows the evolution of conductance 
in a system with isotropic hopping at $t_z/m_{2z}=0.5$ 
with varying film thicknesses
under the truncated boundary condition
as in Fig.~\ref{G-m}(a). % and Fig.~\ref{t20}(a).
 The behavior of the conductance in Fig.~\ref{G-m_zebra} 
looks much more irregular 
compared with that in Fig.~\ref{G-m}(a).
%As a whole,
 One can observe that 
the evolution of the even/odd striped pattern
as in Fig.~\ref{t20}(a) remains
in the downward-pointing triangular region around $m_0/m_2=-2$.
 Outside this region,
new plateaus are observed
beyond the parameter range where the film becomes a QSHI in the 2D limit ($N_z=1$),
in contrast to Fig.~\ref{t20}(a).
 We note that the emergent plateaus arise
even with an {\it even} number of layers.
 The breakdown of the even/odd feature occurs in 
specific parameter regimes
that are specified by the analytics in effective 2D systems
[see Eq.~\eqn{ellipse}]
%and is consistent with the analytics in effective 2D systems.
%this erratic behavior 
and is expected to come from 
emergent 2D topological phases.
%as discussed in the following section.

%%%%% Z2 map  %%%%%
\section{The $\mathbb Z_2$ index map} \label{sec:Z2map}

 In contrast to the case of a 3D bulk TI, Eq.~(\ref{H_bulk}),
a TI thin film, Eq.~(\ref{H_film}),
can be regarded as an effective 2D system,
to which 2D topological classification is applied.
 As a consequence,
the phase diagram of the thin film acquires a richer structure, 
which depends on the thickness of the film.
%In the reminder of the section, we highlight topological aspects of the TI thin film.

%%% Z2 indices %%%
\subsection{Two-dimensional $\mathbb Z_2$ indices for thin film}
 A thin film, or a system with slab geometry of finite thickness,
is regarded as an effective 2D system.
 Actually it does not have to be thin,
as long as the area of the film is infinite,
because the features associated with the motion in the direction of the thickness of the film
can be regarded as those of an internal degree of freedom.
 Here, we take this view and consider Eq.~(\ref{H_film})
to be an effective 2D bulk Hamiltonian for this quasi-2D system.

 Topological classification of gapped free fermionic insulating phases,
which physically mean
either an insulating or a superconducting phase,
is here applied to our thin-film model.
 The periodic table of TIs and superconductors
\cite{Schnyder,Kitaev09,Ryu}
for the 10 universality classes
\cite{Altland97}
%or the so-called ten-fold way,
tells us which type of topological invariant can be used to characterize our system,
i.e., whether $\mathbb Z$- or $\mathbb Z_2$-type.
 The underlying bulk 3D TI of our thin-film model 
is prescribed by Eq.~(\ref{H_bulk}) in the clean limit
and belongs to class DIII.
 However, since we are more concerned about features robust against disorder,
we here assume that on-site disorder is implicit
in our thin-film model and, also, in the parent bulk TI.
 In the presence of potential disorder,
the bulk TI falls into class AII; 
so does its thin-film analog 
except in the limit of $N_z=1$ or $t_z=0$,
where the film Hamiltonian is block diagonalized and each block belongs to class A.
 The topological classification applicable to a system of symmetry class AII
is $\mathbb Z_2$ in both two and three dimensions.

%% parity %%
 The $\mathbb Z_2$-topological 
index $\nu^{\rm 2D}_0$ 
is protected by
the time-reversal symmetry
$\Theta H\_{film}({\bm k}) \Theta^{-1} = H\_{film}(-{\bm k})$,
where $\Theta=-{\rm i} \beta \alpha_x \alpha_z K$ is the time-reversal operator and 
$K$ is the complex conjugate operator.
 Under space inversion symmetry,
the $\mathbb Z_2$-invariants
can be written
in terms of the parity eigenvalues $\xi_{2m}(\Gamma_i)=\pm 1$ of the occupied bands
($m=1,2,\cdots,N$) 
at time-reversal invariant momenta (TRIM) $\Gamma_i$,
\cite{FuKane_inv}
as
%___ nu ___
\begin{align}
   (-1)^{\nu^{\rm 2D}_0} &= \prod_{m=1}^{N_z} \xi_{2m}(0,0) \xi_{2m}(\pi,0) \xi_{2m}(0,\pi) \xi_{2m}(\pi,\pi).
 \label{eqn:nu_0}
\end{align}
%---
 Here, it is implied that in the Brillouin zone of the 2D square lattice, 
there are four time-reversal invariant momenta, 
$\{\Gamma_i\}=\{(0,0), (\pi,0), (0,\pi), (\pi,\pi)\}$,
and $2N_z$ of the total $4N_z$ bands are occupied,
as long as the Fermi energy is in the gap.

 Using eigenstates $|n,\bm k\rangle$ found explicitly for the model Hamiltonian, 
Eq.~(\ref{H_film}),
we can determine the parity eigenvalue $\xi_{2m}(\Gamma_i)$
such that
%___ xi ___
\begin{align}
  P |2m,\bm k=\Gamma_i\rangle
  = \xi_{2m}(\Gamma_i) |2m,\bm k=\Gamma_i\rangle.
\label{xi}
\end{align}
%---
 Note that 
the $2m$th and $(2m-1)$th eigenstates are (Kramers) degenerate, 
and the Kramers doublet share the same parity eigenvalue, 
$\xi_{2m}(\Gamma_i)=\xi_{2m-1}(\Gamma_i)$.
 Collecting all such parity eigenvalues for (half of) the occupied states,
we evaluated the 2D $\mathbb Z_2$-indices given in Eqs.~\eqn{nu_0}--\eqn{nu_y}.
 Repeating this procedure for different mass parameters and for different number of layers
allows for the determination of the phase diagram of a TI thin film.
 In the evaluation of this formula, it is naturally essential to identify
an appropriate inversion operator $P$
that 
satisfies 
$P H\_{film}({\bm k}) P^{-1} = H\_{film}(-{\bm k})$, 
$P\psi_z({\bm k})=\psi_{N_z-z+1}(-{\bm k})$, and 
$[P,\Theta]=0$.
 In the case of a TI nanofilm, $P$ is given as
%___ P ___
\begin{align}
 P=
 \begin{pmatrix}
    &         & 1\\[-1.5mm]
    & \iddots &  \\[-1.0mm]
  1 &         & 
 \end{pmatrix}
 \otimes \beta.
\label{P}
\end{align}
%---
 Clearly, the first part of Eq.~(\ref{P}) plays the role of reversing the layer indices,
while the second part reverses the momentum parallel to the film.

 In parallel with Eq.~\eqn{nu_0},
it is useful to define the 2D weak indices ${\bm \nu}^{\rm 2D}=(\nu^{\rm 2D}_x\, \nu^{\rm 2D}_y)$,
as the 2D analog of the more standard 3D weak indices:
%___ 2Dweak ___
\begin{align}
   (-1)^{\nu^{\rm 2D}_x} &= \prod_{m=1}^{N_z} \xi_{2m}(\pi,0) \xi_{2m}(\pi,\pi), \label{eqn:nu_x}\\
   (-1)^{\nu^{\rm 2D}_y} &= \prod_{m=1}^{N_z} \xi_{2m}(0,\pi) \xi_{2m}(\pi,\pi). \label{eqn:nu_y}
%\label{eqn:2Dweak}
\end{align}
%---
 Similar 2D weak indices have been introduced, e.g., in Ref.~\onlinecite{YIFH}.
 These weak indices are associated with
chiral symmetry, which manifests on the line $k_x, k_y = 0,\pi$.
 Different topological sectors are specified by
the three topological indices,
$\nu^{\rm 2D}_0;(\nu^{\rm 2D}_x\, \nu^{\rm 2D}_y)$.

%%% Z2 map %%%
\subsection{$\mathbb{Z}_2$-index map and emergent topological phases}

 In Fig.~\ref{t05}(a),
these 2D $\mathbb{Z}_2$ topological indices are calculated,
varying the gap parameter $m_0/m_2$ and the thickness $N_z$,
then tabled:
this is the ``$\mathbb{Z}_2$-index map''.
 In the map
the effective QSHI phase, or a 2D TI phase with 
$\nu^{\rm 2D}_0;(\nu^{\rm 2D}_x\, \nu^{\rm 2D}_y) = 1;(0\,0)$ or $1;(1\,1)$,
is shown in yellow,
the trivial (OI) phase with $0;(0\,0)$ in blue,
and
the 2D analog of the 3D WTI phases 
[$0;(1\,1)$ phase in Fig.~\ref{t05}(a)] in red,
to facilitate comparison with the corresponding conductance map
[Fig.~\ref{t05}(b)]. %, see Sec.~\ref{sec:condmap} for the detailed description of conductance maps].
 Note that
the two maps show a reasonably good agreement
for small $N_z$.
 In Fig.~\ref{t05}(a),
the expected even/odd feature of 2D TI states
can be seen only for the region around $m_0/m_2=-2$;
outside this region,
an effective QSHI, or 2D TI (with $\nu^{\rm 2D}_0=1$) phase (yellow region),
appears also at an even number of layers.
 Yellow bands of the QSHI phase are
extended to a broader range of parameters
on the side of smaller $|m_0/m_2|$,
where the single layer film goes into the OI phase;
this implies that a 2D TI phase emerges when OI films are stacked.
 As a whole,
an increasing number of topologically different sectors emerges
as the number of layers is increased.

 The evolution of the TI and OI phases as a function of $N_z$ 
demonstrated here
represents a feature specific to the case of
a relatively small $t_z$; in Fig.~\ref{t05} this is chosen as $t_z/m_{2z}=0.5$,
while
the conductance maps shown in Fig.~\ref{t20}
are calculated in the regime of a relatively large $t_z$ ($t_z/m_{2z}=2$).
 All of these features do not necessarily manifest in
the case of large $t_z$.
 The global structure of the phase diagram changes at $|t_z/m_{2z}|=1$ 
(see Sec.~\ref{sec:zeros} and the Appendix).
 In the regime of small $t_z$ as in Fig.~\ref{t05}
the surface-state wave function shows a damped oscillatory behavior
into the bulk, while
once the hopping exceeds the threshold value $t_z/m_{2z}=1$,
the surface-state wave function is overdamping. 
 Indeed, in the regime of large $t_z$, $t_z/m_{2z}>1$,
the $\mathbb{Z}_2$-index map becomes very simple
and we find the calculated $\mathbb{Z}_2$-index map not insightful enough to be explicitly shown.
 By numerical estimations,
we have verified that
the $\mathbb{Z}_2$ index $\nu^{\rm 2D}_0$ in this regime
takes the values
%___ chi,eps ___
\begin{align}
   \nu^{\rm 2D}_0=
   \left\{
      \begin{array}{cc}
         1 \; & (-5<{m_0\over m_2}<-1  \;\;{\rm and} \;\; N_z={\rm odd}), \\
         0 \; & ({\rm otherwise}),
      \end{array}
   \right.
 \label{eqn:nu_evenodd}
\end{align}
%---
except for $m_0/m_2=-5,\,-3,\,-1$ with $N_z =$ odd, where $\mathbb{Z}_2$ index is not well defined 
(see Sec.~\ref{sec:zero_general}).

%%% phase boundary %%%
\subsection{Zeros of the thin film Hamiltonian: A simplified consideration} \label{sec:zeros}

 In the case of the 3D bulk TI of Eq.~(\ref{H_bulk}),
when the system evolves from a certain TI phase to another, say, 
from an STI to a WTI,
with varying model parameters,
the bulk energy gap collapses in between.
 The locus of such a gap closing
gives the phase boundaries of the
3D system mentioned in Sec.~\ref{sec:intro}.
 In the case of a TI thin film,
the locus of the gap closing of Eq.~(\ref{H_film})
gives the phase boundary of the 2D system.
 Although
the $\mathbb{Z}_2$-index map
in the regime of small $t_z$
[Fig.~\ref{t05}(a)]
looks complicated compared with
the simple behavior in the large-$t_z$ regime,
the location of the phase boundaries 
can be specified systematically in both regimes.

 Let us start from the simple case of vanishing $t_z$.
 In the limit $t_z\rightarrow 0$,
the 2D bulk Hamiltonian,
Eq.~(\ref{H_film}),
can be trivially diagonalized,
since the inter-layer part of $H_z$ reduces %(apart from a prefactor $-m_{2z}\beta/2$)
to a matrix of the form
%___ 1 ___
\begin{align}
   \begin{pmatrix}
    0 & 1      &         &   \\[-2.0mm]
    1 & \ddots & \ddots  &   \\[-1.5mm]
      & \ddots & \ddots  & 1 \\[-0.0mm]
      &        & 1       & 0
   \end{pmatrix}.
 \label{A=0}
\end{align}
%---
 This is diagonalized by a unitary matrix $U$,
and naturally the operation of $U$ and $U^{-1}$ leaves $H_{xy}$ invariant.
 Physically, 
each component $\psi_z$ ($z=1,2,...,N_z$) of the state vector
applied to $H_z$ (and also to $H_{\rm film}$ as a whole)
is an amplitude of the wave function in the corresponding layer $z$.
 Here we choose $\psi_z$ such that
$\psi_z = \sin k_z z$ 
in order to be compatible with the boundary condition
$\psi_0=\psi_{N_z+1}=0$. 
 This imposes a constraint on the allowed value of $k_z$
such that
%___ kz ___
\begin{align}
   k_z^{(n)} = {n\pi\over N_z +1},
 \label{eqn:kz}
\end{align}
%---
where $n=1,2,\cdots,N_z$.
 The corresponding eigenvalues of the matrix, 
Eq.~(\ref{A=0})
is $2\cos k_z^{(n)}$, which takes the values:
$0$ for $N_z=1$,
$\pm 1$ for $N_z=2$,
0 and $\pm \sqrt{2}$ for $N_z=3$, etc.
 Noting that
%___ UHU(tz=0) ___
\begin{align}
   U^{-1}& H_{\rm film}(k_x,k_y) U \nonumber \\
   = &\,  {\rm diag}
          \[m_{\rm 2D}^{(k_x,k_y)} - m_{2z}\cos k_z^{(n=1,...,N_z)} \] \otimes \beta \nonumber \\
     &+ 1_{N_z}\otimes  \sum_{\mu = x, y}t_\mu \sin{k_\mu} \alpha_\mu,
\end{align}
%---%yoshimura 0437
the energies are easily evaluated as
%___ E(tz=0) ___
\begin{align}
   & E_n(k_x,k_y) \nonumber\\
   & =  \pm \sqrt{\( {m_{\rm 2D}^{(k_x,k_y)}\! -\! m_{2z}\cos k_z^{(n)} } \)^2 
       \!+\! \sum_{\mu = x, y}t_\mu^2 \sin^2{k_\mu} } .
\end{align}
%---%yoshimura 0437
 Note that at the phase boundary, 
$E_n$ must vanish at either of the time-reversal symmetric points.
 At the four symmetric points, 
the 2D effective Dirac mass takes the following values:
%___ E(tz=0) ___
\begin{align}
 m_{\rm 2D}^{(0  ,0  )} & = m_0+m_{2z},
 \nonumber \\
 m_{\rm 2D}^{(\pi,0  )} & = m_0+m_{2z}+2m_{2x},
 \nonumber \\
 m_{\rm 2D}^{(0  ,\pi)} & = m_0+m_{2z}+2m_{2y},
 \nonumber \\
 m_{\rm 2D}^{(\pi,\pi)} & = m_0+m_{2z}+2m_{2x}+2m_{2y}.
\label{eqn:m_2D_TRIM}
\end{align}
%---

 In the trivial 2D limit, i.e., the case of $N_z=1$,
the phase transition occurs 
when $m_{\rm 2D}^{(k_x,k_y)}=0$, 
i.e., with four values of $m_0=m_{0{\rm c}(N_z=1)}^{(k_x,k_y)}$:
%___ E(tz=0) ___
\begin{align}
 m_{0{\rm c}(N_z=1)}^{(0  ,0  )} & = -m_{2z},
 \nonumber \\
 m_{0{\rm c}(N_z=1)}^{(\pi,0  )} & = -m_{2z}-2m_{2x},
 \nonumber \\
 m_{0{\rm c}(N_z=1)}^{(0  ,\pi)} & = -m_{2z}-2m_{2y},
 \nonumber \\
 m_{0{\rm c}(N_z=1)}^{(\pi,\pi)} & = -m_{2z}-2m_{2x}-2m_{2y}.
\label{eqn:N=1}
\end{align}
%---
 If Wilson terms are isotropic in the $(x,y)$ plane
and denoted
%___ E(m_para/=0) ___
\begin{align}
 m_{2x}=m_{2y}=m_{2\parallel}\neq 0,\quad
 m_{2z}=m_{2\perp},
\end{align}
%---
%and 
%$m_{2z}=m_{2\perp}$
%for the consistency of notation,
then
two adjacent parameter regimes (let us call them ``QSH regimes''),
%___ QSH2(tz=0) ___
\begin{align}
 -{m_{2\perp} \over m_{2\parallel}}-2 \ <\  {m_0 \over m_{2\parallel}}& \ <\  -{m_{2\perp} \over m_{2\parallel}},
 \label{QSH1}
 \\
 -{m_{2\perp} \over m_{2\parallel}}-4 \ <\  {m_0 \over m_{2\parallel}}& \ <\  -{m_{2\perp} \over m_{2\parallel}}-2,
 \label{QSH2}
\end{align}
%---
separated by a ``degenerate'' phase boundary,
$m_0/m_{2\parallel}=-m_{2\perp}/m_{2\parallel}-2$,
turn out to be
twin QSH regimes with a nontrivial spin Chern number, $\pm 1$.
 The two QSH regimes belong to a sector of different spin Chern number
\cite{BHZ}
but they are identical in the $\mathbb Z_2$ classification;
both correspond to
the $\nu^{\rm 2D}_0=1$ phase.
\cite{KaneMele_Z2}
 On the phase boundary $m_0/m_{2\parallel}=-m_{2\perp}/m_{2\parallel}-2$,
two degenerate Dirac cones appear at two distinct time-reversal momenta,
$(k_x, k_y)=(\pi,0)$ and $(0,\pi)$, in the 2D Brillouin zone.

 In the case of a bilayer TI nanofilm with $N_z=2$,
the phase boundaries given in the single-layer limit
as in Eq.~\eqn{N=1} split into two,
%___ m_c(0,0)(Nz=2,tz=0) ___
\begin{align}
 m_{0{\rm c} (N_z=2)}^{\Gamma_i} = m_{0{\rm c}(N_z=1)}^{\Gamma_i} \pm {m_{2\perp} \over 2},
\label{N=2}
\end{align}
%---
as a result of the hopping between layers,
described by $H_z$ 
as given in Eq.~(\ref{H_z}).
 The amount of the shift leading to the splitting
is determined by the eigenvalue of $H_z$;
in the present case of $N_z=2$, 
it is simply given as $\pm 1$ in units of the prefactor $-m_{2\perp}/2$
of the corresponding hopping term
[see Eq.~(\ref{A=0}) and the text following the equation].
 This leads to the pair of transition points 
around $m_{2\perp}/m_{2\parallel}=-1$ in the $N_z=2$ row in Fig.~\ref{t05}(a),
though in Fig.~\ref{t05}(a) $t_z$ is not 0 but finite, $t_z/m_{2\perp}=0.5$.
 In the range sandwiched between two pairs of degenerate gapless points at $(\pi,0)$ and $(0,\pi)$,
%___ 2Dweak ___
\begin{align}
 -{m_{2\perp} \over m_{2\parallel}} -2 - {1 \over 2}{m_{2\perp} \over 2m_{2\parallel}} 
 \,<\,  {m_0 \over m_{2\parallel}}& 
 \,<\,  -{m_{2\perp} \over m_{2\parallel}} -2 + {1 \over 2} {m_{2\perp} \over 2m_{2\parallel}},
 \label{}
\end{align}
%---
a ``{\it weak} QSHI phase''
[see Eqs.~\eqn{nu_x} and \eqn{nu_y}]
with two helical edge modes appears:
the $G=4$ region indicated in red
[see the $N_z=2$ row in Fig.~\ref{t05}(b)].
%2D weak indices
%for characterizing such weak QSHI phases
%has been introduced in Eqs.~\eqn{nu_x} and \eqn{nu_y}.

In the case of a trilayer film ($N_z=3$), 
similarly to the bilayer case,
$m_{0{\rm c}(N_z=1)}^{\Gamma_i}$
in Eq.~\eqn{N=1} split into three as
%___ m_c(Nz=3,tz=0) ___
\begin{align}
   m_{0{\rm c}(N_z=3)}^{\Gamma_i} = m_{0{\rm c}(N_z=1)}^{\Gamma_i}
   +{m_{2\perp}\over 2}\times
   \left\{\begin{array}{l}
     0,\\
     \pm\sqrt{2}.
   \end{array}\right. 
\label{N=3}
\end{align}
%---
 Note that, in contrast to the case of an even number of layers,
in the case of an odd number of layers,
$H_z$ always bears a null eigenvalue.

%%% analytic %%%
\subsection{Zeros of the thin-film Hamiltonian: A generic case} \label{sec:zero_general}

 By generalizing the discussion above
to nonzero $t_z$
 (see the Appendix),
we obtain $4N_z$ gap closing points,
%___ gap closing points (analytic) ___
\begin{align}
   m_{0{\rm c}(N_z)}^{(0  ,0  )} =& - m_{2z} +\lambda_n, \nonumber \\
   m_{0{\rm c}(N_z)}^{(\pi,0  )} =& - m_{2z} - 2m_{2x}             + \lambda_n, \nonumber \\
   m_{0{\rm c}(N_z)}^{(0  ,\pi)} =& - m_{2z} - 2m_{2y}             + \lambda_n, \nonumber \\
   m_{0{\rm c}(N_z)}^{(\pi,\pi)} =& - m_{2z} - 2m_{2x} - 2m_{2y}   + \lambda_n,
 \label{eqn:m_0c}
\end{align}
%---
with
%___ eigenvalues ___
\begin{align}
 \lambda_n &= \sqrt{m_{2z}^2 - t_z^2} \cos \( {{n\pi} \over {N_z+1}} \), \quad(n=1,...,N_z).
\end{align}
%---
The gapped states of the 2D square lattice are classified into $2^4=16$ topologically distinct phases\cite{YIFH}
and the gapless points correspond to a boundary 
between those topologically distinct phases (not necessarily a TI-OI transition).

%%%%%%%%%%%%%%%%%%%%%%%%%%
\begin{figure}[tbp]
\begin{tabular}{c}
\includegraphics[width=85mm, bb =0 0 506 232]{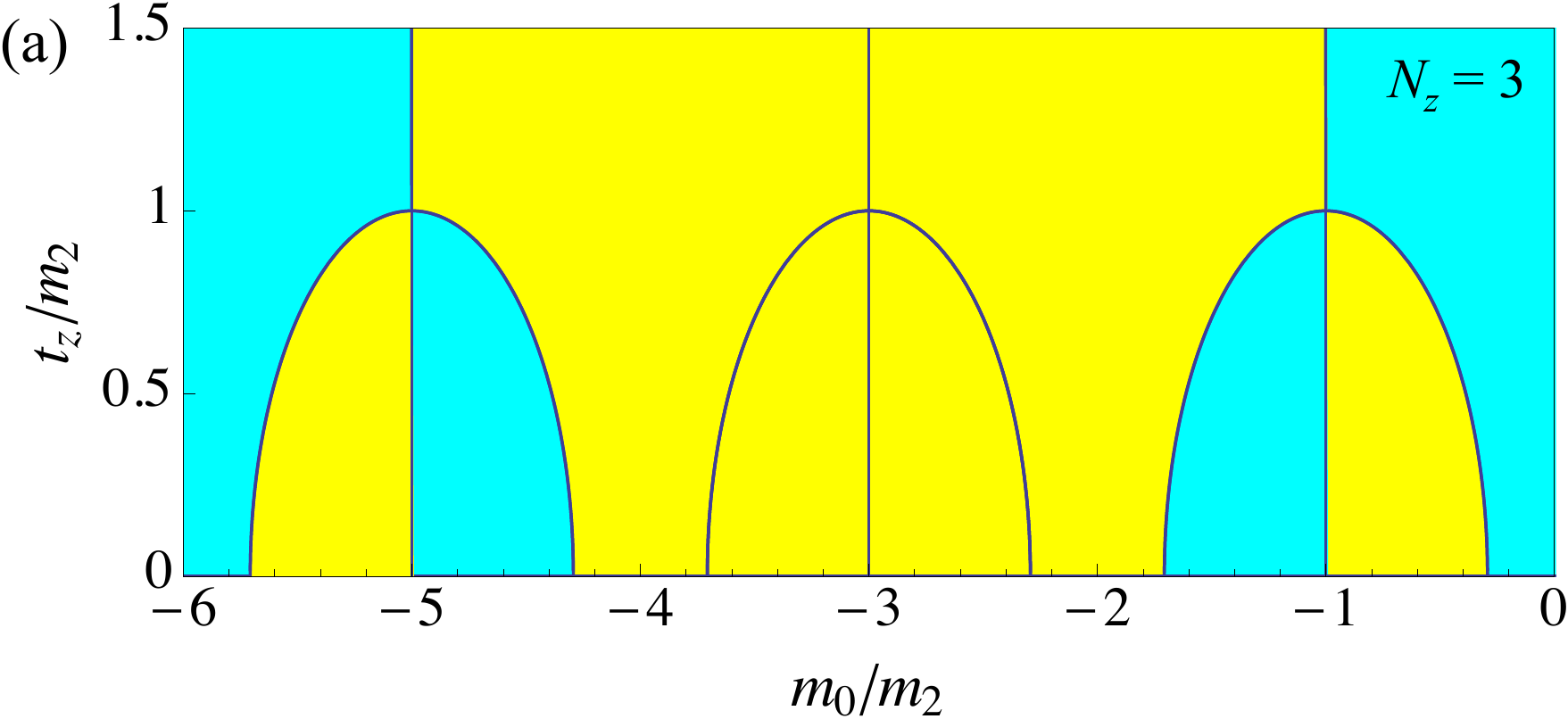}
\\
\includegraphics[width=85mm, bb =0 0 506 232]{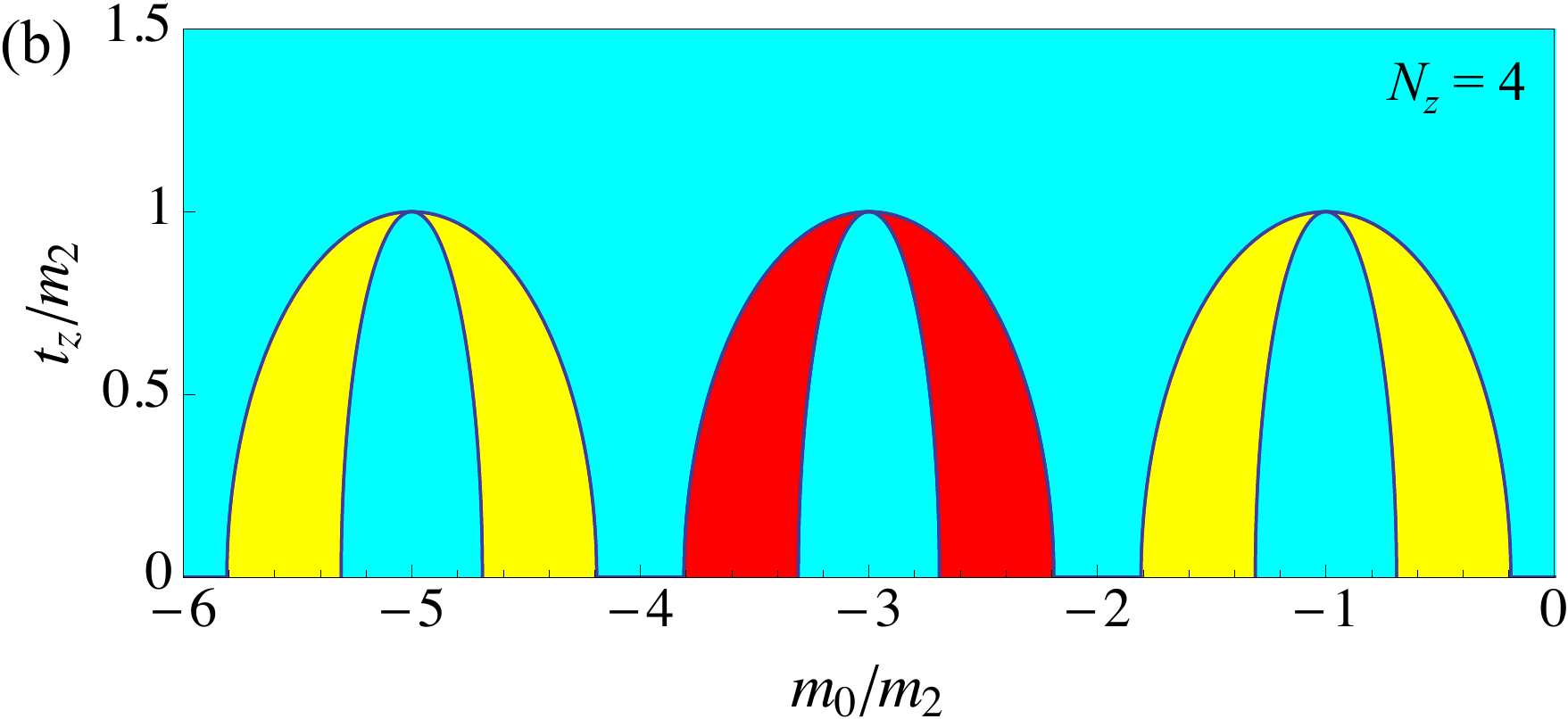}
\end{tabular}
\vspace{-2mm}
\caption{(Color online)
 Evolution of the 2D $\mathbb Z_2$ indices of isotropic film ($m_{2x}=m_{2y}=m_{2z}=m_{2}$) 
as a function of $t_z/m_{2z}$.
 The film thickness is (a) $N_z=3$ and (b) $N_z=4$.
 Solid lines are the zeros of the 2D bulk energy gap
specified by Eq.~\eqn{m_0c},
derived in the Appendix,
and 
they correspond to 2D topological phase boundaries.
 Colors have the same meaning as in Fig.~\ref{t05},
and the intersection of the phase diagram at $t_z/m_{2z}=0.5$
corresponds to the $N_z=3$ or $N_z=4$ row in Fig.~\ref{t05}(a).
}
\label{PD_Az}
\end{figure}
%%%%%%%%%%%%%%%%%%%%%%%
%while the discrete points are zeros of the thin film Hamiltonian (\ref{H_film})
%determined by its direct numerical diagonalization;

 Examples ($N_z=3,4$) 
of the evolution of the phase diagram with respect to $t_z$ are shown in Fig.~\ref{PD_Az}.
 As shown in there,
the behavior of the phase diagram changes drastically at $|t_z/m_{2z}|=1$,
which is implied in Eq.~\eqn{m_0c}.
 In the regime of large $t_z$, $|t_z/m_{2z}|>1$,
only four of the $m_{0\rm c}$ ($\cos k_z^{(n)}=0$, for $n={{N_z+1}\over 2}$) are 
real,
and the 2D topological phase boundaries are the same as those of $N_z=1$, 
Eq.~\eqn{N=1}, 
for $N_z=$ odd
and always trivial ($\cos k_z^{(n)}\neq 0$, for all $n$) for $N_z=$ even.
 As a consequence,
only the even-odd feature of the QSH regime
[see Eqs.~(\ref{QSH1}) and (\ref{QSH2})]
is persistent.
 In the regime of small $t_z$, $|t_z/m_{2z}|<1$,
an increasing number of gapless points (or topologically distinct sectors) emerges 
as an increasing number of layers is stacked,
in the range 
%___  ___
\begin{align}
     m_{0{\rm c}(N_z=1)}^{\Gamma_i}-\sqrt{m_{2z}^2 - t_z^2}
   < m_0
   < m_{0{\rm c}(N_z=1)}^{\Gamma_i}+\sqrt{m_{2z}^2 - t_z^2}.
 \label{eqn:ellipse}
\end{align}
%---
 In contrast to the continuous deformation of the Hamiltonian 
(e.g., switching off the inter-layer coupling\cite{mayu1}),
stacking the layer can lead to emergent topological phenomena;
topologically trivial 2D systems can be driven into 2D TIs by just stacking layers
[see, for example, around $m_0/m_2=-0.5$ in Fig.~\ref{t05}(a)].

 We mention that 
even in the limit $N_z \to \infty$,
the 2D topological phase boundaries
do not converge to the 3D phase boundaries 
(in the case of Figs.~\ref{t20} and \ref{t05}, 
the 3D WTI/STI boundary is located at $m_0/m_2=-2$ 
and the 3D STI/OI boundary at $m_0/m_2=0$).
 This is not an inconsistency, however, 
since the 2D and 3D $\mathbb{Z}_2$ invariants are different by definition.
 The bulk of a TI thin film
can always, 
i.e., irrespective of the number of layers,
be regarded as an effective 2D system and 
characterized by 2D $\mathbb{Z}_2$ invariants,
while 3D $\mathbb{Z}_2$ invariants 
cannot be defined in the film Hamiltonian, 
Eq.~(\ref{H_film}).
 Although we have revealed a rich structure of emergent topological phases in TI thin films as 2D TIs,
we found it difficult 
to reach the 3D limit and 
to investigate the dimensional crossover behavior in our thin-film model.

%%% emergent %%%
%\subsection{Effective 2D description and emergent topological phases}

 Mapping onto the conductance map Fig.~\ref{t05}(b),
one may intuitively expect that 
its erratic behavior 
comes from the emergent topological phases 
as a 2D TI.
 At $N_z=2$ in Fig.~\ref{t05}(b),
the conductance jumps to a quantized value $G=2$
in the range $-1.4\lesssim m_0/m_2\lesssim -0.6$,
while
for a large $t_z$
it is {\it a priori} supposed to vanish [conductance ``valley''; see Fig.~\ref{t20}(a)].
 Similarly, on the other side, around $m_0/m_2\lesssim -2.6$
it is again quantized at $G=4$.
 The border between the original conductance valley
and the new plateau is located at
the zeros of the film Hamiltonian, 
Eq.~(\ref{H_film}),
$m_0/m_2 = -1 \pm \sqrt{3}/4,\, -3 + \sqrt{3}/4$
[see Eq.~\eqn{m_0c}].
%$m_0 = -m_{\rm 2D}^{\Gamma_i} \pm \sqrt{3}/4.$
 On the contrary, at $N_z=3$ the conductance 
is {\it a priori} quantized at $G=2$, 
while this original plateau is replaced by a new valley
around $-1.6\lesssim m_0/m_2$, then reappears, forming a new plateau in the range
$-0.95\lesssim m_0/m_2\lesssim -0.45$.
 The borders between the two plateaus and the new valley 
are again given by 
Eq.~\eqn{m_0c},
$m_0/m_2 = -1,\, -1 \pm \sqrt{6}/4.$
%$m_0 = -m_{\rm 2D}^{\Gamma_i},\, -m_{\rm 2D}^{\Gamma_i} \pm \sqrt{6}/4.$

 What are the origins of these new valleys and plateaus?
 In the parameter regime, Eq.~\eqn{ellipse},
the surface wave function is oscillatory,
\cite{mayu1}
and
the precise behavior of the damped oscillation
continuously varies as a function of the mass parameter.
 Thus, with changing $m_0/m_2$, it happens that
$\psi_{z=N_z}$ vanishes at a critical value of $m_0/m_2$,
then changes sign;
correspondingly,
the hybridization gap vanishes, then gets 
{\it inverted}.
 In the regime of this inverted hybridization gap,
edge states appear just as in the usual QSHI phase.

%%%%%%%%%%%%%%%%%%%%%%%%%%
\begin{figure*}[htbp]
\begin{tabular}{c}
\includegraphics[width=175mm, bb =0 0 741 220]{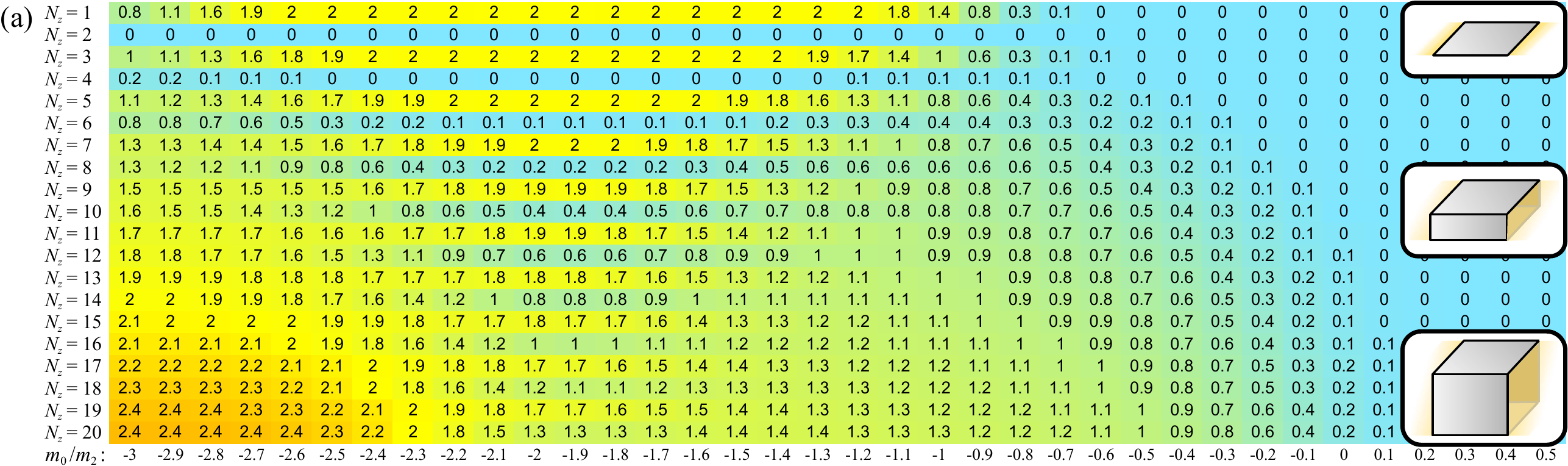}
\\
\includegraphics[width=175mm, bb =0 0 741 220]{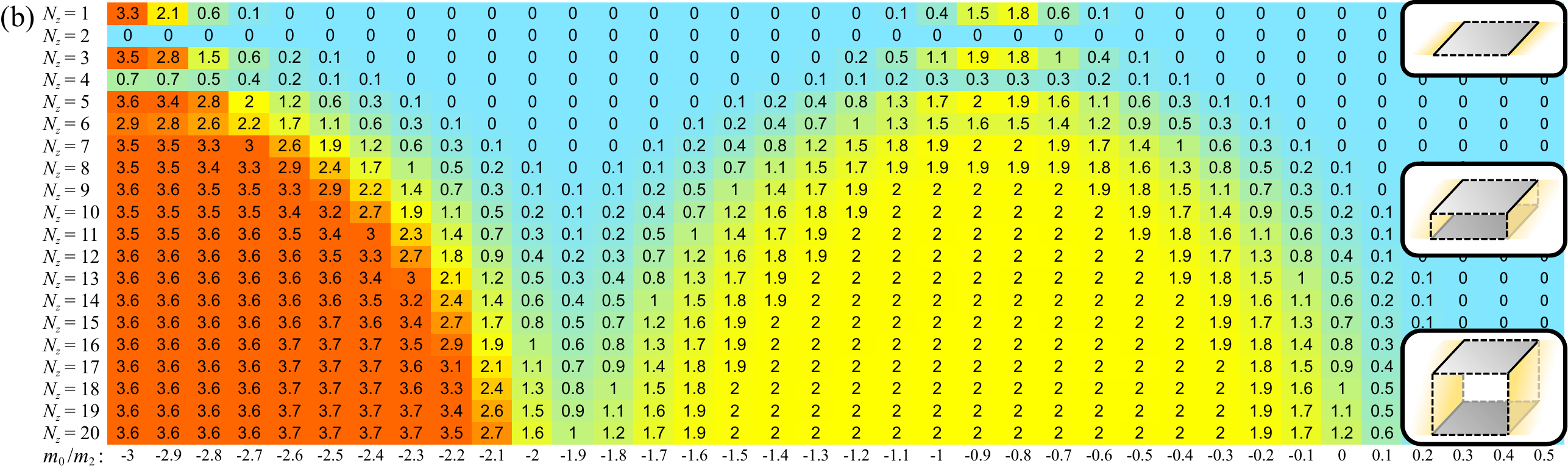}
\end{tabular}
\vspace{-2mm}
\caption{(Color online)
 Conductance maps
with (a) truncated and (b) periodic boundaries on the $y$ sides.
 Parameters are the same as in Fig.~\ref{t20}(a) and \ref{t20}(b)
but, here, in the presence of potential disorder ($W=3$).
 The average of the conductance over $1000$ samples is plotted.
}
\label{cm_W3}
\end{figure*}
%%%%%%%%%%%%%%%%%%%%%%%
%%%%%%%%%%%%%%%%%%%%%%%%%%
\begin{figure*}[htbp]
\begin{tabular}{c}
\hspace{2.5mm}
\includegraphics[width=171mm, bb =0 0 483 147]{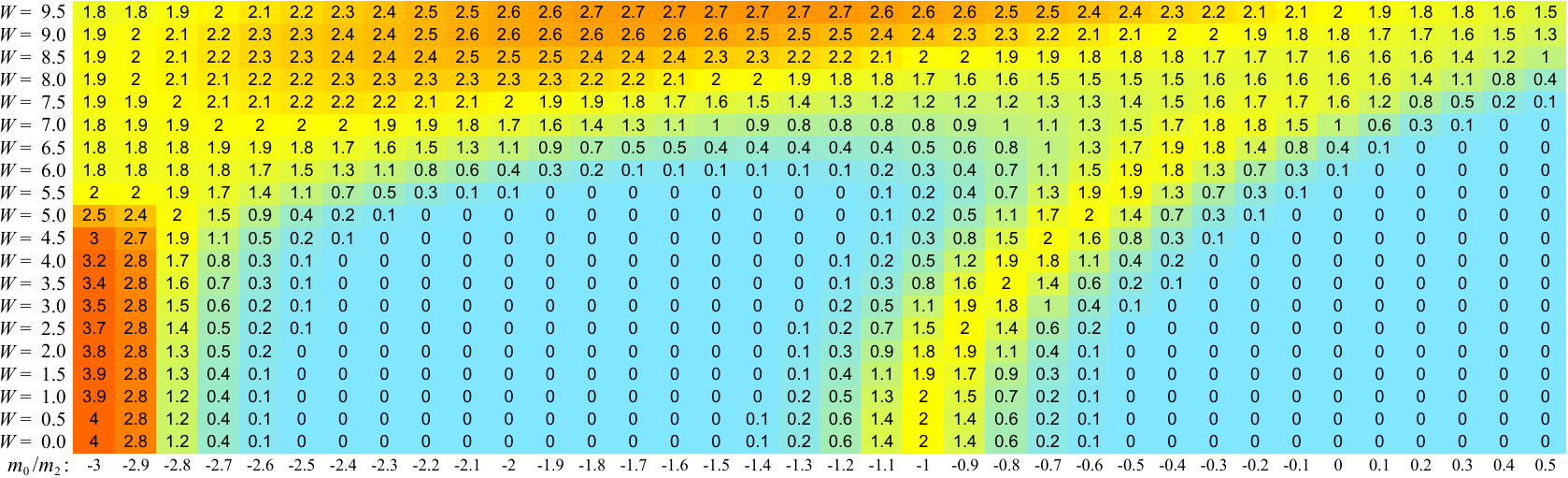}
\end{tabular}
\vspace{-2mm}
\caption{(Color online)
 Conductance map for a disordered slab ($N_z=3)$, with varying disorder strength $W$.
 Other settings are the same as for Fig.~\ref{cm_W3}(b).
}
\label{PD_N3}
\end{figure*}
%%%%%%%%%%%%%%%%%%%%%%%

%%%%%  SEC. disordered  %%%%%
\section{Effects of disorder} \label{sec:disorder}

 The behavior of the conductance in the presence of disorder is an interesting issue to study.
 Now we examine how robust the features are 
that we have so far established to be
characteristic of the conductance maps in the clean limit.
 We introduce the on-site random potential of the box distribution $v_{\bf x}=[-{W\over 2},{W\over 2}]$ into Eq.~\eqn{H_real},
%___ eqn:H_disorder ___
\begin{align}
   H &=  \sum_{\bf x} \sum_{\mu} \(\frac{{\rm i}t_\mu}{2} c^{\dag}_{{\bf x}+{\bf e}_\mu} \alpha_{\mu} c_{\bf x}
                                      -{m_{2\mu}\over 2}  c^{\dag}_{{\bf x}+{\bf e}_\mu} \beta c_{\bf x} + \rm{h.c.}\)  \nonumber \\
%         & + (m_0+ m_{2x}+ m_{2y}+ m_{2z})\sum_{\bf x} c^{\dag}_{\bf x} \beta c_{\bf x}
         & + \(m_0+ \sum_{\mu} m_{2\mu}\)\sum_{\bf x} c^{\dag}_{\bf x} \beta c_{\bf x}
         + \sum_{\bf x} v_{\bf x} c^{\dag}_{\bf x} 1_{4} c_{\bf x},
\label{eqn:H_disorder}
\end{align}
%___
where the summation of $\mu$ runs over $x,y,z$.
 Figures~\ref{cm_W3}(a) and \ref{cm_W3}(b)
are conductance maps under two boundary conditions
analogous to those in Fig.~\ref{t20},
but in the presence of on-site disorder $W=3$.
 One can see that 
the addition of disorder leads to 
effects associated with the so-called ``topological Anderson insulators'',
i.e., the emergence of nontrivial topological characters 
upon the addition of disorder.
\cite{TAI1,TAI2}
 This is a feature due to renormalization of the Dirac mass, which 
eventually changes its sign
when the strength of disorder attains a certain critical value.
\cite{TAI2,ai_JPSJ}
 In Fig.~\ref{cm_W3}
this appears as a shift of phase boundaries 
in comparison with those in Fig.~\ref{t20}.
 However, one can also confirm there that, apart from this shift,
the addition of disorder does not lead to a dramatic effect
on the features of conductance maps in the clean limit
that we have discussed so far.

 To highlight the topological Anderson insulator feature
we have also established a 
conductance map %phase diagram 
(see Fig.~\ref{PD_N3}),
showing the evolution of 
distinct topological phases and their boundaries
as a function of the disorder strength at a fixed number of layers:
$N_z=3$.
 In the region of relatively small $W$ 
($\lesssim 7$)
in Fig.~\ref{PD_N3}
one can recognize two insulating regimes
separated by a narrow region of a quantized conductance of $G=2$
starting at $m_0/m_2=-1$ and $W=0$.
 The left insulating region centered at $m_0/m_2=-2$ at $W=0$,
is the 2D TI (QSHI) phase, while the right region corresponds to
the OI phase.
 The phase boundary
with a quantized conductance of $G=2$
is the 2D version of the Dirac semimetal line studied in Ref.~\onlinecite{KOIH}.
 At an $m_0/m_2$ slightly larger than the location of this Dirac semimetal line,
a system in the OI phase is converted to a 
QSHI
upon the 
addition of disorder (the topological Anderson insulator behavior).
 At $W\gtrsim 7$
the two insulating phases are both overwhelmed by 
a diffusive metal phase, a region of large and nonquantized conductance.
 For the strongly disordered regime $W\gtrsim 15$, 
the system re-enters the OI (Anderson insulator) phase.
 These features 
are most reminiscent of 
a similar phase diagram obtained earlier for a purely 2D system of the same class AII symmetry,
\cite{ai_JPSJ}
based on calculation of the localization length.
 In the structure of the phase diagram shown in Fig.~\ref{PD_N3},
the existence of 
a diffusive metal phase 
is characteristic of systems of class AII symmetry.
 Here,
the initial 2D model, the $N_z=1$ case of Eq.~(\ref{H_film}) with on-site disorder,
belongs to class A,
while stacking more than two layers converts the system to class AII.
 In contrast to this,
only in 
the addition of Rashba-type spin-orbit coupling in the 2D
setup, the system is converted to a model of class AII symmetry
in Ref.~\onlinecite{ai_JPSJ}.
The setup of the present nanofilm construction is much simpler,
and it is an alternative way to realize a class AII QSH system
without
assuming an $s_z$ nonconserving term.

%%%%% conclusion %%%%%
\section{Conclusions} \label{sec:conclusion}

 We have studied the dimensional crossover of TI nanofilms,
focusing on the two types of perfect conduction: surface conduction and edge conduction.
 Our new conjecture
on the two conduction regimes
serves as a guideline to interpolate 2D and 3D limits.
 Based on this conjecture,
the simple picture 
%of WTI based on the observation:~
``(QSH)$^{N_z} \simeq$ WTI''
has been questioned. %from two respects.
 Stacking QSH insulators, i.e., ``(QSH)$^{N_z}$'',
does not necessarily lead to the expected WTI (001) state, 
but to a WTI with other indices or even to an STI as well.
 For the weak interlayer coupling case $m_{2z} \ll m_{2\parallel}$,
2D QSH and 3D WTI (001) states are indeed linked by an edge conduction regime,
but upon approaching the strong interlayer coupling limit $m_{2z} = m_{2\parallel}$,
the edge-conduction regimes in a large number of layers are replaced by 
surface-conduction regimes corresponding to STI or WTI (111) states.

 In the regime of small $t_z$,
transport properties in TI nanofilms become rather erratic,
reflecting the complicated phase diagram of 2D $\mathbb Z_2$ indices.
 In the second part of the paper,
we have revealed the topological properties of TI thin films
(systems with finite thickness and infinite area)
as effective 2D systems.
 We have investigated the 2D $\mathbb Z_2$ phase diagram
by direct calculation of the $\mathbb Z_2$ indices and 
analytical derivation of the locus of the phase boundary for an arbitrary number of layers.
 The phase diagram implies that
a simple even/odd feature
with respect to the number of stacked QSH layers sometimes fails.
 Furthermore, stacked layers of nontopological insulators can enter an emergent 2D topological phase.
 From the bulk point of view,
such an emergent topological phase can also be explained by the inversion of the hybridization gap
of pseudogapless surface states.
%from the viewpoint of the 3D topological insulator with 

%The advantage of our method is that
 The strong point of our analysis is that
by studying the 2D and 3D topological signatures on an equal footing
for different film thicknesses,
we could follow precisely
how the system evolves %(in dimension)
from the 2D limit to the 3D limit.
%from one limit to the other (from 2D to 3D).
 Last but not least,
the employed method is also valid in the presence of disorder.
%we can find both 2D and 3D topological signature
%not only in the dimensional crossover region,
%but also in disordered systems.
 We have shown that our results are stable against disorder unless the bulk band gap closes.
\cite{KOI,KOIH}
 They are also robust against a change in $E$ as long as $E$ is inside the bulk gap,
although in this paper, the calculation has been performed for $E=0$.
 Here, we assumed a square film geometry ($L_x = L_y$).
 In the case of a rectangular film ($L_x \gg L_y$), perfect conduction along the $x$ direction will be easier to observe.
%when the current direction is along the long side of the rectangle.
%even for $T/=0$.
%Film: edge-conduction.
%surface-conduction: L > coherence length.

%%%%%  ack  %%%%%
\begin{acknowledgments}
 Y.Y. and K.I. are indebted to Takahiro Fukui 
for helping us to understand and implement different ways of calculating the $\mathbb Z_2$ indices.
 The authors also acknowledge Yoichi Ando and Yositake Takane for useful comments on this work.
 Y.Y. is supported by JSPS Research Fellowship for Young Scientists.
 The work was supported by JSPS KAKENHI Grant Nos. 15H03700, 15J06436, 15K05131, and 24000013.
\end{acknowledgments}

%%%%% Appendix %%%%%
\appendix
%%% Analytic solution %%
\section{Determination of the phase boundaries in thin films} \label{sec:analytic}
 Diagonalization of the thin-film Hamiltonian, Eq.~(\ref{H_film}), is not straightforward in general.
 The gap closing %of the ``bulk Hamiltonian'' (\ref{H_film})
relevant to the topological phase transition
occurs only at the four inversion and time-reversal symmetric points $\Gamma_i$:
$(0,0), (\pi,0),(0,\pi),(\pi,\pi)$.
At these points, the Hamiltonian is reduced to
%___ H_film_Gamma ___
\begin{align}
 H_{\rm film}(\Gamma_i) =
  &\ 
   1_{N_z}\!\otimes \!
   \bigl(
     m_{\rm 2D}^{(\Gamma_i)} \beta
   \bigr) 
   +H_z ,
\label{eqn:H_film_Gamma}
\end{align}
%---
where $H_z$ is Eq.~(\ref{H_z}) and $m_{\rm 2D}^{(\Gamma_i)}$ are Eq.~\eqn{m_2D_TRIM}.
Applying $\beta$ in Eq.~\eqn{H_film_Gamma} from the left-hand side,
%___ beta H_film_Gamma ___
\begin{align}
 \beta H_{\rm film}(\Gamma_i) =
  &\ 
     m_{\rm 2D}^{(\Gamma_i)}\ 1_{4N_z}
   -\begin{pmatrix}
    0      & \eta_+ &           &        \\[-2.0mm]
    \eta_- & \ddots & \ddots    &        \\[-1.0mm]
           & \ddots & \ddots    & \eta_+ \\[-0.0mm]
           &        & \eta_-    & 0
   \end{pmatrix},
\label{eqn:betaH}
\end{align}
%---
where
%___ eta ___
\begin{align}
   \eta_{\pm} &=      {m_{2z}\over 2} 1_4  
                  \pm {\rm i}{t_{z}\over 2} \beta \alpha_z.
\end{align}
%---
Note that $\eta_{\pm}$ are nonunitary matrices and 
%can be written in the form
%%___ eta ___
%\begin{align}
%   \eta_{\pm} &=  \sqrt{\kappa} \exp(\pm {\rm i} \beta\alpha_z) \nonumber\\
%              &=  \sqrt{\kappa} \[ \cosh(\beta\alpha_z) \pm {\rm i} \sinh(\beta\alpha_z) \],
%\end{align}
%%---
%___ eta eta ___
\begin{align}
   \eta_{-}\eta_{+} = \eta_{+}\eta_{-} = \kappa\ 1_4,
 \label{eqn:etaeta}
\end{align}
%---
where
%___ kappa ___
\begin{align}
   \kappa &= \({m_{2z}\over 2}\)^2 - \Bigl({t_{z} \over 2}\Bigr)^2.
\end{align}
%---
 To find the gapless points, we assume that
$H\_{film}{\bm u}=0$ for an appropriate state vector $\bm u$.
 Then  Eq.~\eqn{betaH} is considered to be a generalized eigenvalue problem for a non-Hermitian matrix:
%___ Hz ___
\begin{align}
   \begin{pmatrix}
    0      & \eta_+ &        &        \\[-2.0mm]
    \eta_- & \ddots & \ddots &        \\[-1.0mm]
           & \ddots & \ddots & \eta_+ \\[-0.0mm]
           &        & \eta_- & 0
   \end{pmatrix}
   {\bm u}
   = m_{\rm 2D}^{(\Gamma_i)} {\bm u}.
\label{eqn:m=eta}
\end{align}
%---
 Here we assume a right eigenvector of the form
%___ eigenvectors ___
\begin{align}
 &{\bm u}_n = C
   \begin{pmatrix}
      \psi_{1,n}\chi_\pm\\
      \psi_{2,n}\chi_\pm\\
      \vdots\\
      \psi_{N_z,n}\chi_\pm
   \end{pmatrix},
 &\psi_{j,n} = c_{j} \sin \bigl( j k_z^{(n)} \bigr)  , 
%     {1\over \sqrt{N}} \left({\eta \over {{1\over 2} \sqrt{m_{2z}^2 - A_z^2}}}\right)^j \sin\left( {{jn\pi} \over {N+1}} \right).
\end{align}
%---
where $C$ is the normalization factor and
$\chi_\pm$
is an eigenspinor of $\eta_{-}$ with eigenvalue 
$\epsilon_\pm \sqrt{\kappa} = {m_{2z} \over 2}\pm {t_{z} \over 2}$,
%___ chi,eps ___
\begin{align}
   {\eta_{-} \over \sqrt{\kappa}} \chi_\pm = \epsilon_\pm \chi_\pm.
\end{align}
%---
(Note $\chi_+$ and $\chi_{-}$ correspond to the top and bottom surface eigenstates, respectively.)
 Then the scalar coefficients $c_j$ follow the relation,
%___ cj ___
\begin{align}
  \lambda_n c_j \sin\bigl( j k_z^{(n)} \bigr) \chi_\pm
   & = c_{j-1} \sin\bigl( (j-1) k_z^{(n)} \bigr) \eta_{-} \chi_\pm \nonumber\\
   & + c_{j+1} \sin\bigl( (j+1) k_z^{(n)} \bigr) \eta_{+} \chi_\pm,
 \label{eqn:cj}
\end{align}
%---
with $\lambda_n$ the eigenvalue of Eq.~\eqn{m=eta}.
 Using Eq.~\eqn{etaeta},
equation~\eqn{cj} can be simplified by applying $\eta_{-}$ from the left hand side, as
%___ cj2 ___
\begin{align}
  \lambda_n c_j \epsilon_\pm \sqrt{\kappa} \sin\bigl( j k_z^{(n)} \bigr)
   & = c_{j-1} \epsilon_\pm^2 \kappa \sin\bigl( (j-1) k_z^{(n)} \bigr) \nonumber\\
   & + c_{j+1} \kappa \sin\bigl( (j+1) k_z^{(n)} \bigr).
 \label{eqn:cj2}
\end{align}
%---
 One can find the solution of Eq.~\eqn{cj2},
%___ c sol ___
\begin{align}
 c_{j} = \epsilon_\pm^{j-1},
\end{align}
%---
with the normalization factor,
%___ C sol ___
\begin{align}
 {1 \over C^2} = \sum_{j=1}^{N_z} \epsilon_\pm^{2(j-1)} \sin^2\bigl( j k_z^{(n)} \bigr),
\end{align}
%---
and the eigenvalues
%___ eigenvalues ___
\begin{align}
 \lambda_n = 2\sqrt{\kappa}\ \cos  k_z^{(n)} . %{{n\pi} \over {N_z+1}} \). %\quad(n=1,...,N_z) .
\end{align}
%---
 Therefore the gap of the film closes at $4N_z$ points [see Eq.~\eqn{m_0c}] on the complex plane,
where $m_{\rm 2D}^{(\Gamma_i)} = \lambda_n$.
%%___ gap closing points (analytic) ___
%\begin{align}
%   m_{0\rm c}^{(0  ,0  )} =& - m_{2z} +\lambda_n, \nonumber \\
%   m_{0\rm c}^{(\pi,0  )} =& - m_{2z} - 2m_{2x}             + \lambda_n, \nonumber \\
%   m_{0\rm c}^{(0  ,\pi)} =& - m_{2z} - 2m_{2y}             + \lambda_n, \nonumber \\
%   m_{0\rm c}^{(\pi,\pi)} =& - m_{2z} - 2m_{2x} - 2m_{2y}   + \lambda_n.
%\label{zeros}
%\end{align}
%%---

We mention that for $t_{z}^2 < m_{2z}^2$, the emergent 2D topological phases appear inside the ellipse in the $(m_0,t_z)$ space (see also Fig.~\ref{PD_Az}),
%___ eigenvalues ___
\begin{align}
   \({{m_0 - m_{0{\rm c}(N_z=1)}^{\Gamma_i}} \over {\cos k_z^{(n=1)} }} \)^2 + t_{z}^2 = m_{2z}^2,
\end{align}
%---
and this corresponds to the parameter region where the damped oscillating behavior of the surface wave function emerges in the semi-infinite model 
[Eq.~(54) in Ref.~\onlinecite{mayu2}].

\bibliography{TI_nanofilms_151127}

\end{document}